
%
%
%
%
%
%
%
%
\def\standardrisposta{s }\def\reducedrisposta{r }
\def\mplarisposta{mpla }\def\zerorisposta{z }
\def\doublerisposta{d }\def\cartarisposta{e }\def\amsrisposta{y }
\newcount\ingrandimento \newcount\sinnota \newcount\dimnota
\newcount\unoduecol \newdimen\collhsize \newdimen\tothsize
\newdimen\fullhsize \newcount\controllorisposta \sinnota=1
\newskip\infralinea  \global\controllorisposta=0
\immediate\write16 { ********  Welcome to PANDA macros (Plain TeX,
AP, 1991) ******** }
\immediate\write16 { You'll have to answer a few questions in
lowercase.}
\message{>  Do you want it in double-page (d), reduced (r)
or standard format (s) ? }\read-1 to\risposta
\message{>  Do you want it in USA A4 (u) or EUROPEAN A4
(e) paper size ? }\read-1 to\srisposta
\message{>  Do you have AMSFonts 2.0 (math) fonts (y/n) ? }
\read-1 to\arisposta
%
%
%
%
\ifx\risposta\standardrisposta \ingrandimento=1200
\message {>> This will come out UNREDUCED << }
\dimnota=2 \unoduecol=1 \global\controllorisposta=1 \fi
\ifx\risposta\reducedrisposta \ingrandimento=1095 \dimnota=1
\unoduecol=1  \global\controllorisposta=1
\message {>> This will come out REDUCED << } \fi
\ifx\risposta\doublerisposta \ingrandimento=1000 \dimnota=2
\unoduecol=2
\message {>> You must print this in
LANDSCAPE orientation << } \global\controllorisposta=1 \fi
\ifx\risposta\mplarisposta \ingrandimento=1000 \dimnota=1
\message {>> Mod. Phys. Lett. A format << }
\unoduecol=1 \global\controllorisposta=1 \fi
\ifx\risposta\zerorisposta \ingrandimento=1000 \dimnota=2
\message {>> Zero Magnification format << }
\unoduecol=1 \global\controllorisposta=1 \fi
\ifnum\controllorisposta=0  \ingrandimento=1200
\message {>>> ERROR IN INPUT, I ASSUME STANDARD
UNREDUCED FORMAT <<< }  \dimnota=2 \unoduecol=1 \fi
\magnification=\ingrandimento
%
%
%
%
\newdimen\eucolumnsize \newdimen\eudoublehsize \newdimen\eudoublevsize
\newdimen\uscolumnsize \newdimen\usdoublehsize \newdimen\usdoublevsize
\newdimen\eusinglehsize \newdimen\eusinglevsize \newdimen\ussinglehsize
\newskip\standardbaselineskip \newdimen\ussinglevsize
\newskip\reducedbaselineskip \newskip\doublebaselineskip
\eucolumnsize=12.0truecm    
\eudoublehsize=25.5truecm   
\eudoublevsize=6.7truein    
\uscolumnsize=4.4truein     
\usdoublehsize=9.4truein    
\usdoublevsize=6.8truein    
\eusinglehsize=6.5truein    
\eusinglevsize=24truecm     
\ussinglehsize=6.5truein    
\ussinglevsize=8.9truein    
\standardbaselineskip=16pt plus.2pt  
\reducedbaselineskip=14pt plus.2pt   
\doublebaselineskip=12pt plus.2pt    
%
%
\def\Portoffset{}
\def\Landoffset{\voffset=-.2truein}
\ifx\risposta\mplarisposta \def\Portoffset{\hoffset=1.8truecm} \fi
%
%
\def\Landspec{}
\tolerance=10000
\parskip=0pt plus2pt  \leftskip=0pt \rightskip=0pt
%
%
\ifx\risposta\standardrisposta \infralinea=\standardbaselineskip \fi
\ifx\risposta\reducedrisposta  \infralinea=\reducedbaselineskip \fi
\ifx\risposta\doublerisposta   \infralinea=\doublebaselineskip \fi
\ifx\risposta\mplarisposta     \infralinea=13pt \fi
\ifx\risposta\zerorisposta     \infralinea=12pt plus.2pt\fi
\ifnum\controllorisposta=0    \infralinea=\standardbaselineskip \fi
\ifx\risposta\doublerisposta   \Landoffset \else \Portoffset \fi
\ifx\risposta\doublerisposta \ifx\srisposta\cartarisposta
\tothsize=\eudoublehsize \collhsize=\eucolumnsize
\vsize=\eudoublevsize  \else  \tothsize=\usdoublehsize
\collhsize=\uscolumnsize \vsize=\usdoublevsize \fi \else
\ifx\srisposta\cartarisposta \tothsize=\eusinglehsize
\vsize=\eusinglevsize \else  \tothsize=\ussinglehsize
\vsize=\ussinglevsize \fi \collhsize=4.4truein \fi
\ifx\risposta\mplarisposta \tothsize=5.0truein
\vsize=7.8truein \collhsize=4.4truein \fi
%
%
%
%
\newcount\contaeuler \newcount\contacyrill \newcount\contaams
\font\ninerm=cmr9  \font\eightrm=cmr8  \font\sixrm=cmr6
\font\ninei=cmmi9  \font\eighti=cmmi8  \font\sixi=cmmi6
\font\ninesy=cmsy9  \font\eightsy=cmsy8  \font\sixsy=cmsy6
\font\ninebf=cmbx9  \font\eightbf=cmbx8  \font\sixbf=cmbx6
\font\ninett=cmtt9  \font\eighttt=cmtt8  \font\nineit=cmti9
\font\eightit=cmti8 \font\ninesl=cmsl9  \font\eightsl=cmsl8
\skewchar\ninei='177 \skewchar\eighti='177 \skewchar\sixi='177
\skewchar\ninesy='60 \skewchar\eightsy='60 \skewchar\sixsy='60
\hyphenchar\ninett=-1 \hyphenchar\eighttt=-1 \hyphenchar\tentt=-1
%
\font\tencmmib=cmmib10  \newfam\cmmibfam  \skewchar\tencmmib='177
\def\bfcal{\cmbsy}                  
\font\tencmbsy=cmbsy10  \newfam\cmbsyfam  \skewchar\tencmbsy='60
\def\scaps{\cmcsc}                 
\font\tencmcsc=cmcsc10  \newfam\cmcscfam
\ifnum\ingrandimento=1095

\font\capsone=cmcsc10 at 10.95pt \font\capstwo=cmcsc10 at 13.145pt

\else

\font\capsone=cmcsc10 at 12pt \font\capstwo=cmcsc10 at 14.4pt
\fi

\def\ttaarr{\bf}		
\def\ppaarr{\sl}		

%
%
%
\newfam\eufmfam \newfam\msamfam \newfam\msbmfam \newfam\eufbfam
\def\Loadeulerfonts{\global\contaeuler=1 \ifx\arisposta\amsrisposta
\font\teneufm=eufm10              
\font\eighteufm=eufm8 \font\nineeufm=eufm9 \font\sixeufm=eufm6
\font\seveneufm=eufm7  \font\fiveeufm=eufm5
\font\teneufb=eufb10              
\font\eighteufb=eufb8 \font\nineeufb=eufb9 \font\sixeufb=eufb6
\font\seveneufb=eufb7  \font\fiveeufb=eufb5
\font\teneurm=eurm10              
\font\eighteurm=eurm8 \font\nineeurm=eurm9
\font\teneurb=eurb10              
\font\eighteurb=eurb8 \font\nineeurb=eurb9
\font\teneusm=eusm10              
\font\eighteusm=eusm8 \font\nineeusm=eusm9
\font\teneusb=eusb10              
\font\eighteusb=eusb8 \font\nineeusb=eusb9
\else \def\eufm{\tt} \def\eufb{\tt} \def\eurm{\tt} \def\eurb{\tt}
\def\eusm{\tt} \def\eusb{\tt}    \fi}
\def\loadeuler{\Loadeulerfonts\tenpoint}
\def\loadamsmath{\global\contaams=1 \ifx\arisposta\amsrisposta
\font\tenmsam=msam10 \font\ninemsam=msam9 \font\eightmsam=msam8
\font\sevenmsam=msam7 \font\sixmsam=msam6 \font\fivemsam=msam5
\font\tenmsbm=msbm10 \font\ninemsbm=msbm9 \font\eightmsbm=msbm8
\font\sevenmsbm=msbm7 \font\sixmsbm=msbm6 \font\fivemsbm=msbm5
\else \def\msbm{\bf} \fi \def\Bbb{\msbm} \def\symbl{\msam} \tenpoint}
\def\loadcyrill{\global\contacyrill=1 \ifx\arisposta\amsrisposta
\font\tenwncyr=wncyr10 \font\ninewncyr=wncyr9 \font\eightwncyr=wncyr8
\font\tenwncyb=wncyr10 \font\ninewncyb=wncyr9 \font\eightwncyb=wncyr8
\font\tenwncyi=wncyr10 \font\ninewncyi=wncyr9 \font\eightwncyi=wncyr8
\else \def\cyrill{\sl} \def\cyrilb{\sl} \def\cyrili{\sl} \fi\tenpoint}
\ifx\arisposta\amsrisposta
\font\sevenex=cmex7               
\font\eightex=cmex8  \font\nineex=cmex9
\font\ninecmmib=cmmib9   \font\eightcmmib=cmmib8
\font\sevencmmib=cmmib7 \font\sixcmmib=cmmib6
\font\fivecmmib=cmmib5   \skewchar\ninecmmib='177
\skewchar\eightcmmib='177  \skewchar\sevencmmib='177
\skewchar\sixcmmib='177   \skewchar\fivecmmib='177
\font\ninecmbsy=cmbsy9    \font\eightcmbsy=cmbsy8
\font\sevencmbsy=cmbsy7  \font\sixcmbsy=cmbsy6
\font\fivecmbsy=cmbsy5   \skewchar\ninecmbsy='60
\skewchar\eightcmbsy='60  \skewchar\sevencmbsy='60
\skewchar\sixcmbsy='60    \skewchar\fivecmbsy='60
\font\ninecmcsc=cmcsc9    \font\eightcmcsc=cmcsc8     \else
\def\cmmib{\fam\cmmibfam\tencmmib}\textfont\cmmibfam=\tencmmib
\scriptfont\cmmibfam=\tencmmib \scriptscriptfont\cmmibfam=\tencmmib
\def\cmbsy{\fam\cmbsyfam\tencmbsy} \textfont\cmbsyfam=\tencmbsy
\scriptfont\cmbsyfam=\tencmbsy \scriptscriptfont\cmbsyfam=\tencmbsy
\scriptfont\cmcscfam=\tencmcsc \scriptscriptfont\cmcscfam=\tencmcsc
\def\cmcsc{\fam\cmcscfam\tencmcsc} \textfont\cmcscfam=\tencmcsc \fi
\catcode`@=11
\newskip\ttglue
\gdef\tenpoint{\def\rm{\fam0\tenrm}
  \textfont0=\tenrm \scriptfont0=\sevenrm \scriptscriptfont0=\fiverm
  \textfont1=\teni \scriptfont1=\seveni \scriptscriptfont1=\fivei
  \textfont2=\tensy \scriptfont2=\sevensy \scriptscriptfont2=\fivesy
  \textfont3=\tenex \scriptfont3=\tenex \scriptscriptfont3=\tenex
  \def\mcal{\fam2 \tensy}  \def\mmit{\fam1 \teni}
  \textfont\itfam=\tenit \def\it{\fam\itfam\tenit}
  \textfont\slfam=\tensl \def\sl{\fam\slfam\tensl}
  \textfont\ttfam=\tentt \scriptfont\ttfam=\eighttt
  \scriptscriptfont\ttfam=\eighttt  \def\tt{\fam\ttfam\tentt}
  \textfont\bffam=\tenbf \scriptfont\bffam=\sevenbf
  \scriptscriptfont\bffam=\fivebf \def\bf{\fam\bffam\tenbf}
     \ifx\arisposta\amsrisposta    \ifnum\contaeuler=1
  \textfont\eufmfam=\teneufm \scriptfont\eufmfam=\seveneufm
  \scriptscriptfont\eufmfam=\fiveeufm \def\eufm{\fam\eufmfam\teneufm}
  \textfont\eufbfam=\teneufb \scriptfont\eufbfam=\seveneufb
  \scriptscriptfont\eufbfam=\fiveeufb \def\eufb{\fam\eufbfam\teneufb}
  \def\eurm{\teneurm} \def\eurb{\teneurb} \def\eusm{\teneusm}
  \def\eusb{\teneusb}    \fi    \ifnum\contaams=1
  \textfont\msamfam=\tenmsam \scriptfont\msamfam=\sevenmsam
  \scriptscriptfont\msamfam=\fivemsam \def\msam{\fam\msamfam\tenmsam}
  \textfont\msbmfam=\tenmsbm \scriptfont\msbmfam=\sevenmsbm
  \scriptscriptfont\msbmfam=\fivemsbm \def\msbm{\fam\msbmfam\tenmsbm}
     \fi      \ifnum\contacyrill=1     \def\cyrill{\tenwncyr}
  \def\cyrilb{\tenwncyb}  \def\cyrili{\tenwncyi}         \fi
  \textfont3=\tenex \scriptfont3=\sevenex \scriptscriptfont3=\sevenex
  \def\cmmib{\fam\cmmibfam\tencmmib} \scriptfont\cmmibfam=\sevencmmib
  \textfont\cmmibfam=\tencmmib  \scriptscriptfont\cmmibfam=\fivecmmib
  \def\cmbsy{\fam\cmbsyfam\tencmbsy} \scriptfont\cmbsyfam=\sevencmbsy
  \textfont\cmbsyfam=\tencmbsy  \scriptscriptfont\cmbsyfam=\fivecmbsy
  \def\cmcsc{\fam\cmcscfam\tencmcsc} \scriptfont\cmcscfam=\eightcmcsc
  \textfont\cmcscfam=\tencmcsc \scriptscriptfont\cmcscfam=\eightcmcsc
     \fi            \tt \ttglue=.5em plus.25em minus.15em
  \normalbaselineskip=12pt
  \setbox\strutbox=\hbox{\vrule height8.5pt depth3.5pt width0pt}
  \let\sc=\eightrm \let\big=\tenbig   \normalbaselines
  \baselineskip=\infralinea  \rm}
\gdef\ninepoint{\def\rm{\fam0\ninerm}
  \textfont0=\ninerm \scriptfont0=\sixrm \scriptscriptfont0=\fiverm
  \textfont1=\ninei \scriptfont1=\sixi \scriptscriptfont1=\fivei
  \textfont2=\ninesy \scriptfont2=\sixsy \scriptscriptfont2=\fivesy
  \textfont3=\tenex \scriptfont3=\tenex \scriptscriptfont3=\tenex
  \def\mcal{\fam2 \ninesy}  \def\mmit{\fam1 \ninei}
  \textfont\itfam=\nineit \def\it{\fam\itfam\nineit}
  \textfont\slfam=\ninesl \def\sl{\fam\slfam\ninesl}
  \textfont\ttfam=\ninett \scriptfont\ttfam=\eighttt
  \scriptscriptfont\ttfam=\eighttt \def\tt{\fam\ttfam\ninett}
  \textfont\bffam=\ninebf \scriptfont\bffam=\sixbf
  \scriptscriptfont\bffam=\fivebf \def\bf{\fam\bffam\ninebf}
     \ifx\arisposta\amsrisposta  \ifnum\contaeuler=1
  \textfont\eufmfam=\nineeufm \scriptfont\eufmfam=\sixeufm
  \scriptscriptfont\eufmfam=\fiveeufm \def\eufm{\fam\eufmfam\nineeufm}
  \textfont\eufbfam=\nineeufb \scriptfont\eufbfam=\sixeufb
  \scriptscriptfont\eufbfam=\fiveeufb \def\eufb{\fam\eufbfam\nineeufb}
  \def\eurm{\nineeurm} \def\eurb{\nineeurb} \def\eusm{\nineeusm}
  \def\eusb{\nineeusb}     \fi   \ifnum\contaams=1
  \textfont\msamfam=\ninemsam \scriptfont\msamfam=\sixmsam
  \scriptscriptfont\msamfam=\fivemsam \def\msam{\fam\msamfam\ninemsam}
  \textfont\msbmfam=\ninemsbm \scriptfont\msbmfam=\sixmsbm
  \scriptscriptfont\msbmfam=\fivemsbm \def\msbm{\fam\msbmfam\ninemsbm}
     \fi       \ifnum\contacyrill=1     \def\cyrill{\ninewncyr}
  \def\cyrilb{\ninewncyb}  \def\cyrili{\ninewncyi}         \fi
  \textfont3=\nineex \scriptfont3=\sevenex \scriptscriptfont3=\sevenex
  \def\cmmib{\fam\cmmibfam\ninecmmib}  \textfont\cmmibfam=\ninecmmib
  \scriptfont\cmmibfam=\sixcmmib \scriptscriptfont\cmmibfam=\fivecmmib
  \def\cmbsy{\fam\cmbsyfam\ninecmbsy}  \textfont\cmbsyfam=\ninecmbsy
  \scriptfont\cmbsyfam=\sixcmbsy \scriptscriptfont\cmbsyfam=\fivecmbsy
  \def\cmcsc{\fam\cmcscfam\ninecmcsc} \scriptfont\cmcscfam=\eightcmcsc
  \textfont\cmcscfam=\ninecmcsc \scriptscriptfont\cmcscfam=\eightcmcsc
     \fi            \tt \ttglue=.5em plus.25em minus.15em
  \normalbaselineskip=11pt
  \setbox\strutbox=\hbox{\vrule height8pt depth3pt width0pt}
  \let\sc=\sevenrm \let\big=\ninebig \normalbaselines\rm}
\gdef\eightpoint{\def\rm{\fam0\eightrm}
  \textfont0=\eightrm \scriptfont0=\sixrm \scriptscriptfont0=\fiverm
  \textfont1=\eighti \scriptfont1=\sixi \scriptscriptfont1=\fivei
  \textfont2=\eightsy \scriptfont2=\sixsy \scriptscriptfont2=\fivesy
  \textfont3=\tenex \scriptfont3=\tenex \scriptscriptfont3=\tenex
  \def\mcal{\fam2 \eightsy}  \def\mmit{\fam1 \eighti}
  \textfont\itfam=\eightit \def\it{\fam\itfam\eightit}
  \textfont\slfam=\eightsl \def\sl{\fam\slfam\eightsl}
  \textfont\ttfam=\eighttt \scriptfont\ttfam=\eighttt
  \scriptscriptfont\ttfam=\eighttt \def\tt{\fam\ttfam\eighttt}
  \textfont\bffam=\eightbf \scriptfont\bffam=\sixbf
  \scriptscriptfont\bffam=\fivebf \def\bf{\fam\bffam\eightbf}
     \ifx\arisposta\amsrisposta   \ifnum\contaeuler=1
  \textfont\eufmfam=\eighteufm \scriptfont\eufmfam=\sixeufm
  \scriptscriptfont\eufmfam=\fiveeufm \def\eufm{\fam\eufmfam\eighteufm}
  \textfont\eufbfam=\eighteufb \scriptfont\eufbfam=\sixeufb
  \scriptscriptfont\eufbfam=\fiveeufb \def\eufb{\fam\eufbfam\eighteufb}
  \def\eurm{\eighteurm} \def\eurb{\eighteurb} \def\eusm{\eighteusm}
  \def\eusb{\eighteusb}       \fi    \ifnum\contaams=1
  \textfont\msamfam=\eightmsam \scriptfont\msamfam=\sixmsam
  \scriptscriptfont\msamfam=\fivemsam \def\msam{\fam\msamfam\eightmsam}
  \textfont\msbmfam=\eightmsbm \scriptfont\msbmfam=\sixmsbm
  \scriptscriptfont\msbmfam=\fivemsbm \def\msbm{\fam\msbmfam\eightmsbm}
     \fi       \ifnum\contacyrill=1     \def\cyrill{\eightwncyr}
  \def\cyrilb{\eightwncyb}  \def\cyrili{\eightwncyi}         \fi
  \textfont3=\eightex \scriptfont3=\sevenex \scriptscriptfont3=\sevenex
  \def\cmmib{\fam\cmmibfam\eightcmmib}  \textfont\cmmibfam=\eightcmmib
  \scriptfont\cmmibfam=\sixcmmib \scriptscriptfont\cmmibfam=\fivecmmib
  \def\cmbsy{\fam\cmbsyfam\eightcmbsy}  \textfont\cmbsyfam=\eightcmbsy
  \scriptfont\cmbsyfam=\sixcmbsy \scriptscriptfont\cmbsyfam=\fivecmbsy
  \def\cmcsc{\fam\cmcscfam\eightcmcsc} \scriptfont\cmcscfam=\eightcmcsc
  \textfont\cmcscfam=\eightcmcsc \scriptscriptfont\cmcscfam=\eightcmcsc
     \fi             \tt \ttglue=.5em plus.25em minus.15em
  \normalbaselineskip=9pt
  \setbox\strutbox=\hbox{\vrule height7pt depth2pt width0pt}
  \let\sc=\sixrm \let\big=\eightbig \normalbaselines\rm }
\gdef\tenbig#1{{\hbox{$\left#1\vbox to8.5pt{}\right.\n@space$}}}
\gdef\ninebig#1{{\hbox{$\textfont0=\tenrm\textfont2=\tensy
   \left#1\vbox to7.25pt{}\right.\n@space$}}}
\gdef\eightbig#1{{\hbox{$\textfont0=\ninerm\textfont2=\ninesy
   \left#1\vbox to6.5pt{}\right.\n@space$}}}
\def\alternativefont#1#2{\ifx\arisposta\amsrisposta \relax \else
\xdef#1{#2} \fi}
\global\contaeuler=0 \global\contacyrill=0 \global\contaams=0
%
%
%
%
\newbox\fotlinebb \newbox\hedlinebb \newbox\leftcolumn
\gdef\makeheadline{\vbox to 0pt{\vskip-22.5pt
     \fullline{\vbox to8.5pt{}\the\headline}\vss}\nointerlineskip}
\gdef\makehedlinebb{\vbox to 0pt{\vskip-22.5pt
     \fullline{\vbox to8.5pt{}\copy\hedlinebb\hfil
     \line{\hfill\the\headline\hfill}}\vss} \nointerlineskip}
\gdef\makefootline{\baselineskip=24pt \fullline{\the\footline}}
\gdef\makefotlinebb{\baselineskip=24pt
    \fullline{\copy\fotlinebb\hfil\line{\hfill\the\footline\hfill}}}
\gdef\doubleformat{\shipout\vbox{\Landspec\makehedlinebb
     \fullline{\box\leftcolumn\hfil\columnbox}\makefotlinebb}
     \advancepageno}
\gdef\columnbox{\leftline{\pagebody}}
\gdef\line#1{\hbox to\hsize{\hskip\leftskip#1\hskip\rightskip}}
\gdef\fullline#1{\hbox to\fullhsize{\hskip\leftskip{#1}%
\hskip\rightskip}}
\gdef\footnote#1{\let\@sf=\empty
         \ifhmode\edef\#sf{\spacefactor=\the\spacefactor}\/\fi
         #1\@sf\vfootnote{#1}}
\gdef\vfootnote#1{\insert\footins\bgroup
         \ifnum\dimnota=1  \eightpoint\fi
         \ifnum\dimnota=2  \ninepoint\fi
         \ifnum\dimnota=0  \tenpoint\fi
         \interlinepenalty=\interfootnotelinepenalty
         \splittopskip=\ht\strutbox
         \splitmaxdepth=\dp\strutbox \floatingpenalty=20000
         \leftskip=\oldssposta \rightskip=\olddsposta
         \spaceskip=0pt \xspaceskip=0pt
         \ifnum\sinnota=0   \textindent{#1}\fi
         \ifnum\sinnota=1   \item{#1}\fi
         \footstrut\futurelet\next\fo@t}
\gdef\fo@t{\ifcat\bgroup\noexpand\next \let\next\f@@t
             \else\let\next\f@t\fi \next}
\gdef\f@@t{\bgroup\aftergroup\@foot\let\next}
\gdef\f@t#1{#1\@foot} \gdef\@foot{\strut\egroup}
\gdef\footstrut{\vbox to\splittopskip{}}
\skip\footins=\bigskipamount
\count\footins=1000  \dimen\footins=8in
\catcode`@=12
\tenpoint
\ifnum\unoduecol=1 \hsize=\tothsize   \fullhsize=\tothsize \fi
\ifnum\unoduecol=2 \hsize=\collhsize  \fullhsize=\tothsize \fi
\global\let\lrcol=L      \ifnum\unoduecol=1
\output{\plainoutput{\ifnum\tipbnota=2 \clearnmbnota\fi}} \fi
\ifnum\unoduecol=2 \output{\if L\lrcol
     \global\setbox\leftcolumn=\columnbox
     \global\setbox\fotlinebb=\line{\hfill\the\footline\hfill}
     \global\setbox\hedlinebb=\line{\hfill\the\headline\hfill}
     \advancepageno  \global\let\lrcol=R
     \else  \doubleformat \global\let\lrcol=L \fi
     \ifnum\outputpenalty>-20000 \else\dosupereject\fi
     \ifnum\tipbnota=2\clearnmbnota\fi }\fi
\def\ifdoublepage{\ifnum\unoduecol=2 }
\gdef\yespagenumbers{\footline={\hss\tenrm\folio\hss}}
\gdef\ciao{ \ifnum\fdefcontre=1 \endfdef\fi
     \par\vfill\supereject \ifnum\unoduecol=2
     \if R\lrcol  \headline={}\nopagenumbers\null\vfill\eject
     \fi\fi \end}

\newskip\olddsposta \newskip\oldssposta
\global\oldssposta=\leftskip \global\olddsposta=\rightskip

\def\filldots{\leaders\hbox to 1em{\hss.\hss}\hfill}
\def\inquadrb#1 {\vbox {\hrule  \hbox{\vrule \vbox {\vskip .2cm
    \hbox {\ #1\ } \vskip .2cm } \vrule  }  \hrule} }
 \def\newline{\hfil\break}
\def\jump{\vskip\baselineskip} \newskip\iinnffrr
\def\sjump{\iinnffrr=\baselineskip
          \divide\iinnffrr by 2 \vskip\iinnffrr}
\def\bjump{\vskip\baselineskip \vskip\baselineskip}
\newcount\nmbnota  \def\clearnmbnota{\global\nmbnota=0}
\newcount\tipbnota \def\letterfootnote{\global\tipbnota=1}

\def\note#1{\global\advance\nmbnota by 1 \ifnum\tipbnota=1
    \footnote{$^{\rm\nttlett}$}{#1} \else {\ifnum\tipbnota=2
    \footnote{$^{\nttsymb}$}{#1}
    \else\footnote{$^{\the\nmbnota}$}{#1}\fi}\fi}
\def\nttlett{\ifcase\nmbnota \or a\or b\or c\or d\or e\or f\or
g\or h\or i\or j\or k\or l\or m\or n\or o\or p\or q\or r\or
s\or t\or u\or v\or w\or y\or x\or z\fi}
\def\nttsymb{\ifcase\nmbnota \or\dag\or\sharp\or\ddag\or\star\or
\natural\or\flat\or\clubsuit\or\diamondsuit\or\heartsuit
\or\spadesuit\fi}   \clearnmbnota
\def\numberfootnote{\global\tipbnota=0} \numberfootnote
\def\setnote#1{\expandafter\xdef\csname#1\endcsname{
\ifnum\tipbnota=1 {\rm\nttlett} \else {\ifnum\tipbnota=2
{\nttsymb} \else \the\nmbnota\fi}\fi} }
\newcount\nbmfig  \def\clearnbmfig{\global\nbmfig=0}
\gdef\figure{\global\advance\nbmfig by 1
      {\rm fig. \the\nbmfig}}   \clearnbmfig
\def\setfig#1{\expandafter\xdef\csname#1\endcsname{fig. \the\nbmfig}}
 \def\endformula{\eqno\numero $$}
 \def\efr{\endformula}
\newcount\frmcount \def\clearfrmcount{\global\frmcount=0}
\def\numero{\global\advance\frmcount by 1   \ifnum\indappcount=0
  {\ifnum\cpcount <1 {\hbox{\rm (\the\frmcount )}}  \else
  {\hbox{\rm (\the\cpcount .\the\frmcount )}} \fi}  \else
  {\hbox{\rm (\applett .\the\frmcount )}} \fi}
\def\nameformula#1{\global\advance\frmcount by 1%
\ifnum\draftnum=0  {\ifnum\indappcount=0%
{\ifnum\cpcount<1\xdef\spzzttrra{(\the\frmcount )}%
\else\xdef\spzzttrra{(\the\cpcount .\the\frmcount )}\fi}%
\else\xdef\spzzttrra{(\applett .\the\frmcount )}\fi}%
\else\xdef\spzzttrra{(#1)}\fi%
\expandafter\xdef\csname#1\endcsname{\spzzttrra}
\eqno \hbox{\rm\spzzttrra} $$}
\def\nfr{\nameformula}    \def\numali{\numero}
\def\nameali#1{\global\advance\frmcount by 1%
\ifnum\draftnum=0  {\ifnum\indappcount=0%
{\ifnum\cpcount<1\xdef\spzzttrra{(\the\frmcount )}%
\else\xdef\spzzttrra{(\the\cpcount .\the\frmcount )}\fi}%
\else\xdef\spzzttrra{(\applett .\the\frmcount )}\fi}%
\else\xdef\spzzttrra{(#1)}\fi%
\expandafter\xdef\csname#1\endcsname{\spzzttrra}
  \hbox{\rm\spzzttrra} }      \clearfrmcount
\newcount\cpcount \def\clearcpcount{\global\cpcount=0}
\newcount\subcpcount \def\clearsubcpcount{\global\subcpcount=0}
\newcount\appcount \def\clearappcount{\global\appcount=0}
\newcount\indappcount \def\clearindappcount{\indappcount=0}
\newcount\sottoparcount 

\def\applett{\ifcase\appcount  \or {A}\or {B}\or {C}\or
{D}\or {E}\or {F}\or {G}\or {H}\or {I}\or {J}\or {K}\or {L}\or
{M}\or {N}\or {O}\or {P}\or {Q}\or {R}\or {S}\or {T}\or {U}\or
{V}\or {W}\or {X}\or {Y}\or {Z}\fi    \ifnum\appcount<0
\immediate\write16 {Panda ERROR - Appendix: counter "appcount"
out of range}\fi  \ifnum\appcount>26  \immediate\write16 {Panda
ERROR - Appendix: counter "appcount" out of range}\fi}
\clearappcount  \clearindappcount \newcount\connttrre
\def\clearconnttrre{\global\connttrre=0} \newcount\countref
\def\clearcountref{\global\countref=0} \clearcountref
\def\chapter#1{\global\advance\cpcount by 1 \clearfrmcount
                 \goodbreak\null\vbox{\jump\nobreak
                 \clearsubcpcount\clearindappcount
                 \itemitem{\ttaarr\the\cpcount .\qquad}{\ttaarr #1}
                 \par\nobreak\jump\sjump}\nobreak}
\def\section#1{\global\advance\subcpcount by 1 \goodbreak\null
               \vbox{\sjump\nobreak\ifnum\indappcount=0
                 {\ifnum\cpcount=0 {\itemitem{\ppaarr
               .\the\subcpcount\quad\enskip\ }{\ppaarr #1}\par} \else
                 {\itemitem{\ppaarr\the\cpcount .\the\subcpcount\quad
                  \enskip\ }{\ppaarr #1} \par}  \fi}
                \else{\itemitem{\ppaarr\applett .\the\subcpcount\quad
                 \enskip\ }{\ppaarr #1}\par}\fi\nobreak\jump}\nobreak}
\clearsubcpcount
\def\appendix#1{\global\advance\appcount by 1 \clearfrmcount
                  \goodbreak\null\vbox{\jump\nobreak
                  \global\advance\indappcount by 1 \clearsubcpcount
          \itemitem{ }{\hskip-40pt\ttaarr #1}
             \nobreak\jump\sjump}\nobreak}
\clearappcount \clearindappcount
\def\references{\goodbreak\null\vbox{\jump\nobreak
   \noindent{\ttaarr References} \nobreak\jump\sjump}\nobreak}

\clearcpcount\clearcountref

\def\setchap#1{\ifnum\indappcount=0{\ifnum\subcpcount=0%
\xdef\spzzttrra{\the\cpcount}%
\else\xdef\spzzttrra{\the\cpcount .\the\subcpcount}\fi}
\else{\ifnum\subcpcount=0 \xdef\spzzttrra{\applett}%
\else\xdef\spzzttrra{\applett .\the\subcpcount}\fi}\fi
\expandafter\xdef\csname#1\endcsname{\spzzttrra}}
\newcount\draftnum \newcount\ppora   \newcount\ppminuti
\global\ppora=\time   \global\ppminuti=\time
\global\divide\ppora by 60  \draftnum=\ppora
\multiply\draftnum by 60    \global\advance\ppminuti by -\draftnum
\def\droggi{\number\day /\number\month /\number\year\ \the\ppora
:\the\ppminuti}     \global\draftnum=0
\def\draftcomment#1{\ifnum\draftnum=0 \relax \else
{\ {\bf ***}\ #1\ {\bf ***}\ }\fi} 
%
%
\catcode`@=11
\gdef\Ref#1{\expandafter\ifx\csname @rrxx@#1\endcsname\relax%
{\global\advance\countref by 1    \ifnum\countref>200
\immediate\write16 {Panda ERROR - Ref: maximum number of references
exceeded}  \expandafter\xdef\csname @rrxx@#1\endcsname{0}\else
\expandafter\xdef\csname @rrxx@#1\endcsname{\the\countref}\fi}\fi
\ifnum\draftnum=0 \csname @rrxx@#1\endcsname \else#1\fi}
\gdef\beginref{\ifnum\draftnum=0  \gdef\Rref{\fairef}
\gdef\endref{\scriviref} \else\relax\fi
\ifx\risposta\mplarisposta \ninepoint \fi
\parskip 2pt plus.2pt \baselineskip=12pt}
\def\Reflab#1{[#1]} \gdef\Rref#1#2{\item{\Reflab{#1}}{#2}}
\gdef\endref{\relax}  \newcount\conttemp
\gdef\fairef#1#2{\expandafter\ifx\csname @rrxx@#1\endcsname\relax
{\global\conttemp=0 \immediate\write16 {Panda ERROR - Ref: reference
[#1] undefined}} \else
{\global\conttemp=\csname @rrxx@#1\endcsname } \fi
\global\advance\conttemp by 50  \global\setbox\conttemp=\hbox{#2} }
\gdef\scriviref{\clearconnttrre\conttemp=50
\loop\ifnum\connttrre<\countref \advance\conttemp by 1
\advance\connttrre by 1
\item{\Reflab{\the\connttrre}}{\unhcopy\conttemp} \repeat}
\clearcountref \clearconnttrre
\catcode`@=12
\ifx\risposta\mplarisposta \def\Reflab#1{#1.} \letterfootnote \fi

\def\slashchar#1{\setbox0=\hbox{$#1$} \dimen0=\wd0
     \setbox1=\hbox{/} \dimen1=\wd1 \ifdim\dimen0>\dimen1
      \rlap{\hbox to \dimen0{\hfil/\hfil}} #1 \else
      \rlap{\hbox to \dimen1{\hfil$#1$\hfil}} / \fi}
\ifx\oldchi\undefined \let\oldchi=\chi
  \def\cchi{{\raise 1pt\hbox{$\oldchi$}}} \let\chi=\cchi \fi
\def\square{\hbox{{$\sqcup$}\llap{$\sqcap$}}}

\def\frac#1#2{{\textstyle{#1 \over #2}}}

\def\half{\ifinner {\scriptstyle {1 \over 2}}\else {1 \over 2} \fi}

\def\simge{\rlap{\raise 2pt \hbox{$>$}}{\lower 2pt \hbox{$\sim$}}}
\def\simle{\rlap{\raise 2pt \hbox{$<$}}{\lower 2pt \hbox{$\sim$}}}

\def\vbig#1#2{{\vbigd@men=#2\divide\vbigd@men by 2%
\hbox{$\left#1\vbox to \vbigd@men{}\right.\n@space$}}}

%
%
\newcount\fdefcontre \newcount\fdefcount \newcount\indcount
\newread\filefdef  \newread\fileftmp  \newwrite\filefdef
\newwrite\fileftmp     \def\strip#1*.A {#1}
\def\futuredef#1{\beginfdef
\expandafter\ifx\csname#1\endcsname\relax%
{\immediate\write\fileftmp {#1*.A}
\immediate\write16 {Panda Warning - fdef: macro "#1" on page
\the\pageno \space undefined}
\ifnum\draftnum=0 \expandafter\xdef\csname#1\endcsname{(?)}
\else \expandafter\xdef\csname#1\endcsname{(#1)} \fi
\global\advance\fdefcount by 1}\fi   \csname#1\endcsname}

\def\beginfdef{\ifnum\fdefcontre=0
\immediate\openin\filefdef \jobname.fdef
\immediate\openout\fileftmp \jobname.ftmp
\global\fdefcontre=1  \ifeof\filefdef \immediate\write16 {Panda
WARNING - fdef: file \jobname.fdef not found, run TeX again}
\else \immediate\read\filefdef to\spzzttrra
\global\advance\fdefcount by \spzzttrra
\indcount=0      \loop\ifnum\indcount<\fdefcount
\advance\indcount by 1   \immediate\read\filefdef to\spezttrra
\immediate\read\filefdef to\sppzttrra
\edef\spzzttrra{\expandafter\strip\spezttrra}
\immediate\write\fileftmp {\spzzttrra *.A}
\expandafter\xdef\csname\spzzttrra\endcsname{\sppzttrra}
\repeat \fi \immediate\closein\filefdef \fi}
\def\endfdef{\immediate\closeout\fileftmp   \ifnum\fdefcount>0
\immediate\openin\fileftmp \jobname.ftmp
\immediate\openout\filefdef \jobname.fdef
\immediate\write\filefdef {\the\fdefcount}   \indcount=0
\loop\ifnum\indcount<\fdefcount    \advance\indcount by 1
\immediate\read\fileftmp to\spezttrra
\edef\spzzttrra{\expandafter\strip\spezttrra}
\immediate\write\filefdef{\spzzttrra *.A}
\edef\spezttrra{\string{\csname\spzzttrra\endcsname\string}}
\iwritel\filefdef{\spezttrra}
\repeat  \immediate\closein\fileftmp \immediate\closeout\filefdef
\immediate\write16 {Panda Warning - fdef: Label(s) may have changed,
re-run TeX to get them right}\fi}
\def\iwritel#1#2{\newlinechar=-1
{\newlinechar=`\ \immediate\write#1{#2}}\newlinechar=-1}
\global\fdefcontre=0 \global\fdefcount=0 \global\indcount=0
%
%
\null
%
%
%
%
\loadamsmath
\loadeuler
%
%
%
%
\mathchardef\subsetneq="3D20
\def\ie{{\it i.e.\/}}

\def\gg{{\>\widehat{g}\>}}

\def\s{{\bf s}}
\def\sw{{{\bf s}_w}}
\def\Heis{{\cal H}[w]}
\def\ad{{\rm ad\;}}
\def\Ker{{\rm Ker}(\ad\Lambda)}
\def\Im{{\rm Im}( \ad\Lambda)}
\def\Pos{{\rm P}_{\geq0[\sw]}}
\def\Neg{{\rm P}_{<0[\sw]}}
\def\W{$\cal W$}
\pageno=0
\nopagenumbers{\baselineskip=12pt
\line{\hfill US-FT/13-94}
\line{\hfill (Revised version)}
\line{\hfill\tt hep-th/9409016}
\line{\hfill August 1994}\bjump
\ifdoublepage \bjump\bjump\bjump\else\jump\vfill\fi
\centerline{\capstwo $\bfcal W$-Algebras from Soliton Equations}
\sjump
\centerline{\capstwo and Heisenberg Subalgebras}
\bjump\bjump\jump
\centerline{{\scaps Carlos
R.~Fern\'andez-Pousa\footnote{$^1$}{\tt
pousa@gaes.usc.es}}, {\scaps Manuel~V.
Gallas\footnote{$^2$}{\tt gallas@gaes.usc.es}},}\jump
\centerline{{\scaps J. Luis Miramontes\footnote{$^3$}{\tt
miramontes@gaes.usc.es}},
and
{\scaps Joaqu\'\i n S\'anchez Guill\'en\footnote{$^4$}{\tt
joaquin@gaes.usc.es}}}
\jump\jump
\centerline{\sl Departamento de F\'\i sica de Part\'\i culas,}
\centerline{\sl Facultad de F\'\i sica,}
\centerline{\sl Universidad de Santiago,}
\centerline{\sl E-15706 Santiago de Compostela, Spain}
\bjump\jump
\ifdoublepage
\vfill
{\noindent
\line{August 1994\hfill}}
\eject\null\vfill\fi
\centerline{\capsone ABSTRACT}\jump

\noindent
We derive sufficient conditions under which the
``second'' Hamiltonian structure of a class of generalized
KdV-hierarchies defines
one of the classical $\cal W$-algebras obtained through Drinfel'd-Sokolov
Hamiltonian reduction. These integrable hierarchies are
associated to the Heisenberg subalgebras of an untwisted
affine Kac-Moody algebra. When the principal Heisenberg subalgebra
is chosen, the well known connection between the
Hamiltonian structure of the generalized
Drinfel'd-Sokolov hierarchies ---the Gel'fand-Dickey
algebras--- and the $\cal W$-algebras associated to the
Casimir invariants of a Lie algebra is recovered. After
carefully discussing the relations between the embeddings
of $A_1=sl(2,{\Bbb C})$ into a simple Lie algebra $g$ and the
elements of the Heisenberg subalgebras of $g^{(1)}$, we
identify the class of $\cal W$-algebras that can be
defined in this way. For $A_n$, this class only includes those  associated to
the embeddings labelled by partitions of the form
$n+1\>=\> k\>(m)\> +\> q\>(1)$ and $n+1\>=\> k\>(m+1)\> +\> k\>(m)\> +\>
q\>(1)$.

\sjump\vfill
\ifdoublepage \else
\noindent
\line{August 1994\hfill}\fi
\eject}
\yespagenumbers\pageno=1
\footline={\hss\tenrm-- \folio\ --\hss}
%

\chapter{{\bf Introduction.}}

The recent interest in \W-algebras is motivated by their wide spectrum of
applications in classical and quantum two-dimensional field theory.
Just to mention some of them, \W-algebras play an important role in the
study of conformal field theories, quantum and topological gravity,
string theories, critical phenomena in statistical mechanics, and even
the quantum Hall effect (see the reviews~[\Ref{REV}]). Moreover, from a
fundamental point of view, they might lead to the formulation of
Lorentz-invariant field theories with symmetries generated
by fields of higher spin.

The \W-algebras are (non-linear) extensions of the Virasoro algebra by
conformal primary fields~[\Ref{ZAM}]. The algebraic structures
related to \W-algebras first appeared in the context of certain
integrable hierarchies of differential equations~[\Ref{GD}]. The
simplest example of this relation is provided by the Korteweg-de Vries
equation (KdV), the archetypical integrable system,
$$
\eqalign{
\partial_{t} u(x,t) & = -{1\over4} u'''(x,t) + {3\over2}u(x,t)u'(x,t)
\cr
& = \{H\>, \>u(x,t)\}_2 \>, \cr}
$$
where $\>'=\partial/\partial x$, the Hamiltonian is $H=\int dx\>
{1\over4} u^2(x,t)$, and the ``second'' Poisson bracket is
$$
\{u(x,t),u(y,t)\}_2 = {1\over2}\delta'''(x-y) -
2u(x,t)\delta'(x-y) - u'(x,t) \delta(x-y)\>,
$$
which is nothing but a classical realization of the Virasoro
algebra. In~[\Ref{DS}], it was shown that generalized equations of the
KdV type can be defined for arbitrary affine Kac-Moody algebras; the
original KdV equation corresponding to $A_{1}^{(1)}$. The ``second''
Hamiltonian structure of these integrable differential equations is
defined by the Gel'fand-Dickey algebras, and it has been
proved~[\Ref{FL}] that they are connected with the \W-algebras
associated to the Casimir invariants of the classical Lie
algebras~[\Ref{BAIS}]. For instance, the second Hamiltonian
structure of the Drinfel'd-Sokolov
$A_{n-1}$ generalized KdV hierarchy is connected with ${\cal W}_n$, the
\W-algebra whose generators have conformal spin
$2,3,\ldots,n$~[\Ref{BAK}].

This early known connection between the
Drinfel'd-Sokolov generalized KdV hierarchies and (certain) \W-algebras
relates them to Lie algebras, and this opens the
possibility of their classification. In fact, among the different
methods used in the literature to construct
\W-algebras, the most effective and systematic one is conceptually
inspired by this connection. It consists in the Hamiltonian reduction of
affine Kac-Moody algebras~[\Ref{REV},\Ref{EMB},\Ref{HAM1},\Ref{HAM2}], and it
is usually called (quantum or classical) Drinfel'd-Sokolov reduction.
With this method, a polynomial extension of the Virasoro algebra can
be associated to each embedding of
$A_1\equiv sl(2,{\Bbb C})$ into a simple finite Lie
algebra~[\Ref{EMB},\Ref{SORB1}]. The Lagrangian field theoretical version of
this procedure involves the reduction of Wess-Zumino-Novikov-Witten models
(WZNW) to, generally non-abelian, Toda field theories, which
are well known two-dimensional integrable field theories. Moreover,
there is convincing evidence to believe that all the \W-algebras can be
constructed in this way~[\Ref{RUELLE}].

Using the Drinfel'd-Sokolov reduction procedure at the classical level,
the \W-algebra associated to the Casimir invariants of the Lie algebra
$g$ is recovered by choosing the principal embedding of $A_1$ into $g$.
The natural question is if it is possible to associate integrable
systems to all the other \W-algebras obtained with other
embeddings. So far, the outcome of all the attempts to answer this
question is the construction of integrable systems associated to some
particular families of
\W-algebras~[\Ref{CASES},\Ref{MATH1},\Ref{MATH2},\Ref{BAKF},
\Ref{VAND},\Ref{BAKDEP},\Ref{FHM},\Ref{BON}],
but still the general structure of this connection has not been developed.

Recently, the Drinfel'd-Sokolov construction has been generalized
showing that it is possible to associate an integrable hierarchy of
differential equations of the KdV type to each particular Heisenberg
subalgebra of an affine Kac-Moody algebra~[\Ref{GEN1},\Ref{GEN2}] (see
also~[\Ref{FHM},\Ref{NIGEL1},\Ref{NIGEL2},\Ref{DFEH}]). The original
Drinfel'd-Sokolov hierarchies are then recovered with the principal
Heisenberg subalgebra. Even more, when the affine Kac-Moody algebra is
simply-laced, it has been proved in ref.~[\Ref{TAU}] that the hierarchy
is one of those constructed by Kac and Wakimoto within the tau-function
approach~[\Ref{KW}]. These integrable hierarchies are bi-Hamiltonian,
and one of their Hamiltonian structures, the ``second'' one, exhibits a
conformal invariance~[\Ref{GEN2}]. Therefore, it is logical to think
that their Hamiltonian structures might be related to the \W-algebras
obtained by Hamiltonian reduction of affine Kac-Moody algebras. It is
the purpose of this paper to establish the precise form of this
connection.

A priori, the approach of refs.[\Ref{GEN1},\Ref{GEN2}] seems quite
different from that of~[\Ref{REV},\Ref{EMB},\Ref{HAM1},\Ref{HAM2}].
In the later, the fundamental objects are the embeddings of
$A_1$ into simple Lie algebras, while, in the former, the construction
is based on the Heisenberg subalgebras of affine Kac-Moody algebras.
Besides, in the Drinfel'd-Sokolov reduction, the
\W-algebras are obtained by reducing a classical Kac-Moody algebra that
is formulated as the Kirillov-Poisson bracket associated to a Lie
algebra $g$, \ie, $g$ is affinized once in a variable $x$. In contrast,
the integrable hierarchies of~[\Ref{GEN1},\Ref{GEN2}] are defined in
terms of Lax operators defined on a loop algebra ${\cal L}(g)=g\otimes
{\Bbb C}[z,z^{-1}]$ ---an affine Kac-Moody algebra in general;
therefore, the phase space consists of functions of a variable $x$
on ${\cal L}(g)$, \ie, $g$ is affinized twice: in $z$ and $x$. However,
notice that the Lie algebra $g$ can be identified with the subalgebra
$g\otimes 1\subset {\cal L}(g)$, which indicates that the phase space
involved in the Drinfel'd-Sokolov reductions is, in fact, a subset of
the phase space where the generalized integrable hierarchies are
defined. We will show how this relation between these two phase spaces
specifies a connection between the embeddings of $A_1$ into $g$ and the
Heisenberg subalgebras of ${\cal L}(g)$.

The paper is organized as follows. In Section~2 we review the
construction of~[\Ref{GEN1},\Ref{GEN2}], where integrable
hierarchies are associated to the non-equivalent Heisenberg subalgebras
of a loop algebra. The definition of these integrable systems involves
the choice of a constant graded element of a Heisenberg subalgebra,
say $\Lambda$. A distinction is made between
``regular'' and ``non-regular'' $\Lambda$'s, leading to type~I and
type~II hierarchies, respectively. Nevertheless, even though it has
been  proved in~[\Ref{FHM},\Ref{DFEH}] that most Heisenberg subalgebras do
not have regular graded elements, type~II hierarchies are not described in
detail in the original references. We have tried to fill this gap in
Section~2.

In the next two sections we analyse the ``second'' Hamiltonian structure
of the integrable hierarchies of~[\Ref{GEN1},\Ref{GEN2}] and we compare
it with the \W-algebras. In Section~3 we discuss the conformal invariance
of the second Poisson bracket and we find the energy-momentum tensor that
generates this conformal transformation on a subspace of the total phase
space. Restricted to this subspace, the Poisson bracket is just the
Kirillov-Poisson bracket associated to a (non-simple, in general) Lie
algebra, say $\gg_0(\s)$. This is the starting point in connecting the
second Poisson bracket algebra with
\W-algebras. The phase space of the integrable hierarchy consists of a
set of equivalence classes under a  group of gauge transformations; we
propose a convenient choice of the gauge slice when $\Lambda$
satisfies a ``non-degeneracy'' condition. In Section~4 we establish the
connection between the second Poisson bracket algebra and the classical
\W-algebras obtained with the Drinfel'd-Sokolov reduction approach.
This connection, explained in theorem~2, gives a precise relation
between the constant graded element $\Lambda$ used in the construction of
the integrable hierarchy and the $J_+$ of the embedding of
$A_1$ into $\gg_0(\s)$ that specifies the corresponding \W-algebra.

With this result, it is possible to investigate which
\W-algebras can be constructed as the second Poisson bracket algebras
of the integrable hierarchies of~[\Ref{GEN1},\Ref{GEN2}]. This is
equivalent to show which nilpotent elements $J_+$ are related to the
graded elements of the Heisenberg subalgebras, and we do it here for the
classical Lie algebras of the
$A_n$ series. Recall that the inequivalent Heisenberg subalgebras of the
untwisted Kac-Moody algebra
$g^{(1)}$ are in one-to-one relation with the conjugacy classes of
the Weyl group of $g$. Therefore, as a previous step, we have summarized
in Section~5 the classification of the $A_1$ subalgebras and the
conjugacy classes of the Weyl group of
$A_n$. In order to establish the required connections between the
elements of all the Heisenberg subalgebras of $A_{n}^{(1)}$ and the
$A_1$ subalgebras of $A_n$. In fact, we derive two kinds of
connections. The first one is based on the properties of the cyclic
element of B.~Kostant~[\Ref{KOST}] and it can be generalized to the
other classical algebras. The second one is based on the
particular structure of the Heisenberg subalgebras of $A_{n}^{(1)}$
and, hence, it is specific of this case. These results are interesting by
themselves, and  we hope that they will be useful, for instance, in the
study of two-dimensional integrable field
theories~[\Ref{UNDER},\Ref{FRENK}]. The class of
\W-algebras that can be constructed as the second Hamiltonian structures
of the integrable hierarchies of~[\Ref{GEN1},\Ref{GEN2}] is obtained in
Section~6. This class is quite limited, even
if type~II hierarchies are included, and the origin of this restriction
is that $J_+$ has to be related to a ``graded'' element
of a Heisenberg subalgebra.

In Section~7 we consider some examples for the purposes of
illustration of our results. Finally, we present our conclusions in
Section~8. The explicit construction of the inequivalent Heisenberg
subalgebras of $A_{n}^{(1)}$, with $n=2,\ldots,5$, is presented in the
Appendix.

\chapter{{\bf The generalized KdV hierarchies.}}

In this section we summarize the construction
of~[\Ref{GEN1},\Ref{GEN2}], and we generalize some of the results
of these works to include the case of the hierarchies of type~II. Unless
otherwise indicated, we follow the conventions of~[\Ref{GEN2}].

The central object is an affine Kac-Moody algebra $\gg$. In this
paper, $\gg$ will be considered as the loop algebra associated to a
finite (simple) Lie algebra $g$,
$\gg=g\otimes {\Bbb C} [z,z^{-1}]\oplus {\Bbb C}\>d$, where $d=z\>
d/dz$, and whose central extension will be ignored (but see
ref.[\Ref{TAU}]). A very important role is played by the Heisenberg
subalgebras of
$\gg$, which are maximal abelian subalgebras of $\gg$~(see~[\Ref{KP}]
for a precise definition). Up to conjugation, they are in one-to-one
correspondence with the conjugacy classes of the Weyl group of
$g$~[\Ref{KP}], and we denote them as $\Heis$, where $[w]$ indicates
the conjugacy class. Associated to $\Heis$ there is a distinguished
gradation $\sw$, with the property that
$\Heis$ is an invariant subspace under the adjoint action
of $d_{\sw}$\note{With the exception of a few cases including the
principal Heisenberg subalgebras, the choice of $\sw$ is not unique,
something that has not been mentioned in~[\Ref{GEN1},\Ref{GEN2}]. We
shall comment on this point later.}. This ensures that one can choose a
basis of $\Heis$ whose elements are graded with respect to $\sw$.

The construction of the integrable hierarchies is based on the
(matrix) Lax equation. The Lax operator is associated to the data
$\{\gg,\Heis,\sw,\s,\Lambda\}$, where \s\ is an additional gradation
such  that $\s\preceq\sw$ with respect to a partial ordering, and
$\Lambda$ is a constant element of $\Heis$ with positive $\sw$-grade
$i$.  Then, the Lax operator is defined as
$$
L= \partial_x + \Lambda + q(x)\>,
\nfr{Lax}
where the potential $q(x)$ is a function of $x$ on the subspace
$$
Q= \gg_{\geq0}(\s)\cap \gg_{<i}(\sw)\>.
\nfr{Potential}
For technical reasons, the potentials are taken to be periodic
functions of $x$~[\Ref{DS}]. Notice that any element $\Lambda\in {\cal
H}[w]$ is semisimple, \ie, $\ad\Lambda$ is  diagonalizable on $\gg$;
then $\gg =
\Ker \oplus \Im$, where $\Ker$ is a subalgebra of $\gg$, and
$[\Ker,\Im]\subseteq
\Im$. A distinction is made between hierarchies of type~I and type~II,
referring to whether the element $\Lambda$ is ``regular'' or not
---regularity implying $\Ker = \Heis$.

The potential $q(x)$ plays the role of the phase space coordinate in
this system. However, there exist symmetries corresponding
to the gauge transformations
$$
L \mapsto \exp ({\rm ad}\> S)\> (L)\>,
\nfr{Gauge}
generated by the functions $S$ of $x$ on the subspace
$$
P = \gg_0(\s) \cap \gg_{<0}(\sw) \>.
\nfr{Gaugegen}
The flows of the integrable hierarchy are defined on the set of gauge
invariant functionals of $q(x)$.
It is straightforward to find a basis for these gauge invariant
functionals. One simply performs a non-singular gauge transformation to
take $q(x)$ to some canonical form $q^{({\rm can})}(x)$; then, the components
of
$q^{({\rm can})}(x)$ and their $x$-derivatives provide the desired
basis.

The flows that define the hierarchies are non-linear partial differential
equations, and it is convenient to demand that they can be written
in the form
$$
{\partial q^{({\rm can})}\over \partial t} = F\left( q^{({\rm can})},
\partial_x
q^{({\rm can})}, \ldots\right)\>,
\nfr{Gaugeflow}
with $F$ being a polynomial functional of the components of $q^{({\rm
can})}$ and  its $x$-derivatives; otherwise one would obtain some sort
of ``rational''  non-linear differential equations. The condition that
the map $\ad \Lambda:P
\mapsto Q$ is injective ensures this polynomiality; it is equivalent
to
$$
\Ker\cap P =\emptyset\>,
\nfr{Degen}
and it  allows the use of the standard Drinfel'd-Sokolov gauge fixing
procedure. Let us recall that, within this procedure,
$q^{({\rm can})}(x)$ is characterized as an element of some complementary
vector space of $[\Lambda,P]$ in $Q$, \ie,  $q^{({\rm can})}(x)\in Q^{({\rm
can})}$ with $Q=Q^{({\rm can})}+ [\Lambda,P]$. The choice of $Q^{({\rm
can})}$ is not unique~(for a very detailed
description of the Drinfel'd-Sokolov gauge fixing in a similar
situation, see~[\Ref{RUELLE}]).

If the hierarchy is of type~I, then $\Ker\cap P =\emptyset$ for
any choice of $\Lambda$ because of the following.

\jump
\noindent
{\bf Lemma 1.} {\it Let $b\in\Heis$ be an element of $\Heis$ that is
graded with respect to the two gradations $\s$ and $\s'$. Then, the
$\s$-grade of  $b$ equals $N_{\s}/ N_{\s'}$ times its $\s'$-grade.}

\sjump
\noindent
{\it Proof.} Recall that the derivation inducing the gradation
$\s$ is
$$
d_\s = N_\s \> z {d\over dz} + H_\s\>.
\nfr{Derivation}
$H_\s$ is an element of the Cartan subalgebra of $g$ such that
$\alpha_j(H_\s)= s_j\in {\Bbb Z}^+$, where $\alpha_1,\ldots,\alpha_{{\rm
rank}(g)}$ is a choice of the simple roots of $g$, and
$$
N_{\s} = \sum_{j=0}^{{\rm rank}(g)} k_j s_j\>,
\efr
with $\{k_0,\ldots,k_{{\rm rank}(g)}\}$ being the Kac labels of $\gg$
(to compare with~[\Ref{GEN2}], notice that $H_\s/N_\s \equiv \delta_\s\cdot
H$).  Therefore,
$$
{d_{\s}\over N_{\s}} - {d_{\s'}\over N_{\s'}} =
{H_{\s}\over N_{\s}} - {H_{\s'}\over N_{\s'}}\equiv \Delta(\s,\s')
\nfr{Relative}
is also an element of the Cartan subalgebra of $g$. If $b\in\Heis$
is graded with respect to the two gradations, then, it has to be stable
under the adjoint action of
$\Delta(\s,\s')$, which means that $[\Delta(\s,\s'), b]= \beta\> b$ for some
rational number $\beta$. Now, taking inner products with the elements
of $\Heis$ and using the invariance of the Killing form
on $\gg$ and the commutativity of $\Heis$, it follows that $\beta=0$.
Consequently,
$$
[d_{\s} ,b]\> =\> {N_{\s}\over N_{\s'}}\, [d_{\s'} ,b]
\nfr{Lemma}
which proves the lemma. $\square$\jump

On the contrary, if the hierarchy is of type~II one has to
restrict the choice of $\Lambda$ in such a way that
$\Ker\cap P =\emptyset$ to ensure the polynomiality of the flows.

The outcome of~[\Ref{GEN1},\Ref{GEN2}] is that there exists an infinite
number of commuting flows on the gauge equivalence classes of $L$.
These flows have the following form. For each constant element $b$ of
the centre of $\Ker$ with non-negative $\sw$-grade there is a flow that
can be defined in many gauge  equivalent
ways parameterized by the generators $\theta(x)\in
P$ of (infinitesimal) gauge transformations\note{The dependence on $x$
should be generalized to include all the
$t_b$'s, but, as usual, we only show explicitly $x$.}:
$$
\eqalign{
{\partial q(x)\over \partial t_{b}^{\theta}} &= \lbrack \Pos\left(\exp (-\ad
V)\>  (b) \right) + \theta, L\rbrack \cr
&= \lbrack -\Neg\left(\exp (-\ad V)\>
(b) \right) + \theta, L\rbrack\>; \cr}
\nfr{FlowsG}
notice that the centre of $\Ker$, ${\rm Cent}\left(\Ker\right)$, is the set
of $x\in \Ker$ such that $[x,y]=0$ for any $y\in \Ker$,
and, obviously, ${\rm Cent}\left(\Ker\right) \subseteq \Heis$. The two
equivalent definitions of the flow on the right-hand-side of \FlowsG,
together with the definition of $Q$ as an intersection of two subspaces of
$\gg$ characterized by the gradations \s\ and $\sw$, ensures that $\partial
q(x)/ \partial t_{b}^{\theta} \in Q$. In \FlowsG, $\Pos$ and $\Neg$ are the
projectors onto $\gg_{\geq0}(\sw)$ and $\gg_{<0}(\sw)$, respectively.

Eqs.\FlowsG\ define a
unique flow on the gauge slice $Q^{({\rm can})}(x)$ because, then, the
function $\theta(x)$ is fixed uniquely by the  condition that $\partial
L^{\rm can}/\partial t_b \in Q^{({\rm can})}$, and  $\theta(x)$ becomes a
polynomial functional of the components of $q(x)$ and its derivatives.
In \FlowsG, $V(x)\in \Im\cap  \gg_{<0}(\sw)$ is the generator of the unique
transformation which takes  the Lax operator to the form
$$
{\cal L}  = \exp (\ad V) (L) = L + [V,L] + {1\over2} [V,[V,L]] + \ldots =
\partial_x + \Lambda + h(x)\>,
\nfr{Diagonal}
where $h(x)$ is a function of $x$ on $\Ker\cap\gg_{<i}(\sw)$ (recall that $i$
is the $\sw$-grade of $\Lambda$). It is important to remark that both $V(x)$
and $h(x)$ are polynomial functionals of the components of $q(x)$ and its
$x$-derivatives\note{In  connection with the integrability of type~II
hierarchies, let us say that it should be possible to associate
a flow not only to the elements of the centre of $\Ker$ but to any constant
element of $\Heis$ also in this case. The difference is that those flows not
associated to the  elements of the centre of $\Ker$ are always
non-polynomial.}

One of the main properties of the flows defined by \FlowsG\ is that
they are Hamiltonian. Even more, they are bi-Hamiltonian when $\s$ is
the  homogeneous gradation~[\Ref{GEN2}]. The two Hamiltonian structures
are defined on the  set of gauge invariant functionals and they
adopt the form of Kirillov-Poisson brackets on the affine  Kac-Moody
algebra
$\gg$. If
$\varphi[q]$ and $\psi[q]$ are two functionals of
$q(x)$, the Poisson bracket defining the ``first'' Hamiltonian
structure is
$$
\lbrace \varphi,\psi \rbrace_1 = -\left( d_q\varphi, z^{-1} \lbrack
d_q\psi , L\rbrack\right) \>,
\nfr{First}
while the bracket for the ``second'' structure is
$$
\lbrace \varphi,\psi \rbrace_2 = \left( d_q\varphi_0, \lbrack
d_q\psi_0 , L\rbrack\right) - \left( d_q\varphi_{<0}, \lbrack
d_q\psi_{<0} , L\rbrack\right)\>,
\nfr{Second}
where the definitions of the inner product $(\cdot\> ,\cdot)$ and of the
functional derivative $d_q \varphi\in \gg$ are standard and can be found
in~[\Ref{GEN2}]. Only those  hierarchies for which $\s=\s_{\rm
hom}=(1,0,\ldots,0)$ admit the two  Hamiltonian structures, whereas the
hierarchies are Hamiltonian only with respect to the second Poisson
bracket in the  general case. In these last two equations, we have used
the following notation: superscripts denote $\sw$-grades, and  subscripts
indicate $\s$-grades, but not just homogeneous grades as in~[\Ref{GEN2}]. For
example, if $a\in\gg$ then $a^j$ is the component of $a$ on the
subspace
$\gg_j(\sw)$, and $a_k$ is its component on $\gg_k(\s)$. We shall
resort to this notation to simplify some expressions.

The Poisson brackets defined above admit
non-trivial centres. In fact, notice that, by construction, the $\s$-grade of
the
components of $q(x)$ is bounded, \ie, there exists an non-negative integer $n$
such that
$$
\Lambda + q(x) \in \bigoplus_{j=0}^{n}\gg_j(\s)\>.
\efr
Then, it follows from \First\ and \Second\ that the components of $q(x)$ whose
$\s$-grade is $n$ are centres of the two Poisson bracket algebras; but they are
not gauge invariant centres in general. In contrast, the existence of gauge
invariant functionals that are centres of the Poisson brackets is
directly related to the components of $h(x)$ in \Diagonal\ with non-negative
$\sw$-grade.

\jump\noindent
{\bf Proposition 1.} {\it The functionals of the form
$\Theta_f = \left(f(x),h(x)\right)$ for $f(x)$ being a function of $x$ on
$$
\Ker^{1-i} \cup {\rm Cent}\left( \bigoplus_{j=1-i}^{k} \Ker^j\right)
\nfr{Centres}
are gauge invariant centres of the first Poisson bracket algebra for $k=0$, and
of the second Poisson bracket algebra for $k=-1$.}
$\square$\jump

This result is the direct generalization of the
Proposition~3.3 of~[\Ref{GEN2}] to include both type~I and type~II
hierarchies; the proof is identical but taking into account that
$\Ker$ is  not abelian in the general case. Moreover, recall that the
functionals $\Theta_f$ for $k=0$ are non-dynamical and that $-x$ can
be identified with the time $t_\Lambda$ only when $\Pos\left(h(x)\right)=0$.
The meaning of the mismatch between the centres of the two Poisson brackets
has been clarified in~[\Ref{NIGEL1}].

The equations of the hierarchy are invariant under the scaling
transformation
$$
x\mapsto \lambda x\>, \qquad t_{b^j}\mapsto \lambda^{j/i} t_{b^j}\>,
\qquad q^k\mapsto \lambda^{k/i -1} q^k\>,
\nfr{Scaling}
where we have introduced a basis $\lbrace e^k\rbrace$ of
$Q$, whose
elements are graded with respect to $\sw$, \ie,
$$
[d_\sw, e^k] = k e^k, \>{\rm\;and\;}\>
q(x) = \sum_{k<i} q^k(x) e^k\>.
\nfr{Grades}
In relation to ${\cal W}$-algebras, the crucial property
is that \Scaling\
can be generalized to a conformal transformation
$$
x\mapsto y(x)\>,\quad q(x;z)\mapsto \>\widetilde{q}\>(y(x);
\widetilde{z})\>,
\nfr{Confone}
with
$$
q(x;z)\> =\> y' \>U[y]\> \biggl(\widetilde{q}\>(y(x);\widetilde{z}\>)\>
\biggr) +  {1 \over i} \left(y''\over y' \right)  H_\sw\>,
\nfr{Conformal}
and this transformation is a Poisson mapping of the second Poisson
bracket.  In \Conformal,
$$
y'={dy(x)\over dx}\>, \qquad\widetilde{z}=(y')^{-{N_{\sw}/ i}}\> z\>,
\qquad U[y] = \left(y'\right)^{- {\ad H_{\sw}/ i}}\>,
$$
and we have explicitly indicated the dependence of the potential on
$z$, the affine parameter defining the loop algebra $\gg$.
The infinitesimal version of \Conformal\ is $y(x) = x + \epsilon(x)$
and
$$
\eqalign{
\delta_\epsilon q(x;z) & = q(x;z) - \>\widetilde{q}\>(x;z) \cr
&= \partial_x\left(
\epsilon(x)\> q(x;z)\right) \> -\> {1\over i}\epsilon'(x)\lbrack
N_{\sw}z{d\over dz} + H_\sw ,  q(x;z)\rbrack + \> {1\over i}
\epsilon''(x)  H_\sw\> \cr
& = \lbrack \partial_x + \Lambda(z) + q(x;z)\>, \epsilon(x)\left(
\Lambda(z) +  q(x;z) \right) \>+ \> {1\over i} \epsilon'(x) H_\sw
\rbrack \cr &\qquad - \>{N_\sw \over i}\> \epsilon'(x)\> \lbrack
z{d\over dz}\>, \>
\Lambda(z) +  q(x;z) \rbrack\>,\cr}
\nfr{Infinitesimal}
which, using \Grades, is just
$$
\eqalign{
\delta_\epsilon q(x;z) & = \sum_{k<i} e^k(z)\> \delta_\epsilon q^k(x)
\cr & = \sum_{k<i} e^k(z)\> \left( \epsilon(x) \partial_x q^k(x) +
\epsilon'(x)\left(1-{k\over i}\right) q^k(x)\right) \>+ \>{1\over i}
\epsilon''(x) H_{\sw}\>.\cr}
\nfr{Infintwo}
This transformation also induces a corresponding conformal transformation on
the space of gauge equivalent classes, which is the phase space of the
hierarchy.

\chapter{{\bf The energy-momentum tensor.}}

The results of~[\Ref{GEN2}] ensure that the gauge invariant
functionals of $q(x)$ close
under the  second Poisson bracket \Second, and they have well defined
properties with respect to the conformal transformation \Conformal. This
suggests a general relationship between the second Poisson bracket
algebra and some (classical) quiral extended conformal algebras whose
generators  would be the components of
$q^{({\rm can})}(x)$, the basis of gauge invariant functionals. This
suggestion is supported by the examples analysed in~[\Ref{GEN2}].
A first step to establish the connection is to show that there is some
gauge invariant functional $T_\epsilon[q]$ that plays the role of the
energy-momentum tensor and generates the infinitesimal conformal
transformation \Infinitesimal. However, let us anticipate that, when
the second Hamiltonian structure is a \W-algebra, it is
defined by a conformal structure generally twisted with respect to the
conformal structure associated to the transformation \Infinitesimal.

The condition that $T_\epsilon[q]$
generates \Infinitesimal\ means that\note{From now on we suppress the
subscript indicating  that the Poisson bracket is the second Poisson bracket,
and the explicit indication of the dependence of
$\Lambda$ and $q$ on $z$.}
$$
\eqalign{
\delta_\epsilon q(x) &= \lbrace T_\epsilon[q]\> ,\> q(x)\rbrace \cr
& = \biggl(\lbrack \partial_x +
 \Lambda + q(x)\>, \> (d_{q} T_\epsilon[q])_0\rbrack\biggr)_0 \, - \,
\biggl(\lbrack \Lambda + q(x)\>, \> (d_{q}
T_\epsilon[q])_{<0}\rbrack\biggr)_{>0}\>.
\cr}
\nfr{InfT}
The generator of the conformal transformation of the components of
$q(x)$ whose $\s$-grade is zero, $q_0(x)$, is easily derived from
eqs.\Infinitesimal\ and \InfT. The result is that
$$
\delta_\epsilon q_0(x) = \lbrace T_\epsilon[q]\> ,\> q(x)\rbrace\>,
\nfr{Inftwo}
with
$$
d_q T_\epsilon = \epsilon(x) \left( \Lambda + q(x)\right)_0\, + \,
{1\over i}\> \epsilon'(x)\>  H_\sw\>,
\nfr{Tinf}
and the energy-momentum tensor is $T_\epsilon[q] = \int dx\, \epsilon(x)\>
T(x)$ with
$$
\eqalign{
T(x) & = {1\over 2}\langle \Lambda + q(x)\>, \> \Lambda + q(x)\rangle_{\gg} \,
-
\, {1\over i} \partial_x\langle  H_\sw\>, \>  \Lambda +
q(x)\rangle_{\gg}\cr
&= {1\over 2}\langle \left(\Lambda + q(x)\right)_0\>, \> \left( \Lambda +
q(x)\right)_0\rangle_{\gg} \, -
\, {1\over i} \partial_x\langle  H_\sw\>, \>  q_0(x)\rangle_{\gg}\> ,\cr}
\nfr{TTT}
where $\langle\cdot\>, \cdot\rangle_{\gg}$ is the Killing form of $\gg$.
In the particular case when $\s$ is the homogeneous gradation, this result
has already been obtained in~[\Ref{NIGEL1}]. Notice that the components of
$d_q T\in \gg_{<0}(\s)$ are not specified, and  we have consistently taken
them to zero. The Poisson bracket of two  energy-momentum tensors can be
calculated and the result is
$$
\lbrace T(x),T(y)\rbrace\> =\> {\langle  H_\sw,  H_\sw\rangle \over i^2}\>
\delta'''(x-y)\> -\> 2\> T(x)\> \delta'(x-y)\> - T'(x)\>
\delta(x-y) \>,
\nfr{Virasoro}
where $\delta^{(n)}(x-y)\equiv \partial_{x}^{(n)}\delta(x-y)$, and, now,
$\langle\cdot\>, \cdot\rangle$ is the Killing form of the finite Lie algebra
$g$. Eq.\Virasoro\ shows that $T(x)$ indeed closes a classical Virasoro
algebra  with respect to the second Poisson bracket whose central extension
is $12
\langle  H_\sw,  H_\sw \rangle /i^2$.

For the other components of $q(x)$, whose $\s$-grade is $>0$, finding the
generator of the conformal transformation  from \Infinitesimal\ and \InfT\ is
not so straightforward. In particular, we already know that some of these
components are centres of the Poisson bracket algebra; so, there will
be no energy-momentum tensor that generates their conformal
transformation. As non-trivial examples involving $q_{>0}(x)$, let us mention
the $A_1$ fractional KdV-hierarchies analysed in~[\Ref{GEN2}], and the
integrable hierarchy associated to $A_5$ that we discuss in Section~7.

The reason why it is easy to derive the generator of the conformal
transformation \Inftwo\ is that the second Poisson bracket restricted
to the functionals of $q_0(x)$ simplifies to
$$
\lbrace \varphi,\psi\rbrace  = \biggl( \lbrack (d_q \varphi)_0, (d_q
\psi)_0 \rbrack\>, \>(\Lambda + q(x))_0
\biggr) - \biggl((d_q \varphi)_0\>, \>\partial_x\left(d_q
\psi\right)_0\biggr)\>,
\nfr{KM}
which is just the Kirillov-Poisson bracket corresponding to the
untwisted affinization in $x$ of the finite Lie algebra $\gg_0(\s)$,
where the affine $\gg_0(\s)$ current has been reduced to
$$
J(x)=(\Lambda + q(x))_0\>;
\nfr{Reduced}
as usual, the last term on the right-hand-side of \KM\ especifies the
central extension. Notice that, because of lemma~1, either $(\Lambda)_0$
vanishes of it is  a nilpotent element of $\gg_0(\s)$. Of
course, the energy-momentum tensor \TTT\ comes from the
Sugawara construction applied to the reduced current $J(x)$ improved by
a term that depends on the gradation
$\sw$.

In the following, instead of considering the general case, we shall
concentrate on the integrable hierarchies for which not only the
condition
\Degen\ is satisfied, but also
$$
{\rm Ker}(\ad(\Lambda)_0) \cap P =\emptyset\>,
\nfr{Degenplus}
which corresponds to the ``non-degeneracy'' condition
of~[\Ref{HAM2},\Ref{SORB1}]; \Degenplus\ is equivalent to
${\rm Ker}(\ad(\Lambda)_0) \cap \gg_{<0}(\sw) =\emptyset$. Therefore, either
$P$ is empty or $(\Lambda)_0\not=0$. If $P=\emptyset$ then the hierarchy is a
generalized mKdV hierarchy with
$\s=\sw$, $(\Lambda)_0=0$, and the second Poisson bracket algebra restricted
to the functionals of $q_0(x)$ is an untwisted Kac-Moody
algebra. In contrast, when $(\Lambda)_0\not=0$, we will show that this second
Poisson bracket algebra is a $\cal W$-algebra. In this case, a convenient
choice of the gauge slice is:

\jump\noindent
{\bf Proposition 2.} {\it Under the condition \Degenplus, the gauge slice
$q^{({\rm can})}(x)$ can be chosen following the Drinfel'd-Sokolov
prescription in such a way that
$$
\left(q^{({\rm can})>0}(x)\right)_0 =0\>,
\nfr{Gaugeone}
and that $q^{({\rm can})\leq0}(x)$ depends only on $q_0(x)$.}

\sjump\noindent
{\it Proof.} First of all, notice that the proposition is obviously true when
$(\Lambda)_0=0$ because $P=\emptyset$; then, \Gaugeone\ is an identity and
$q^{({\rm can})\leq0}(x)= q^{({\rm can)}}_0(x) = q_0(x)$.
Therefore, we shall concentrate on the case of $(\Lambda)_0\not=0$.
Consider the non-singular gauge transformation that takes $q(x)$ to the
gauge slice
$q^{({\rm can})}(x)$
$$
q^{({\rm can})}(x) = \exp\left(\ad \widehat{S}\right) \biggl(\partial_x +
\Lambda +q(x)\biggr) - \partial_x -\Lambda\>,
\nfr{Gaugetrans}
with $\widehat{S}$ a polynomial functional of $q(x)$ and its $x$-derivatives.
According to their $\s$-grade, and taking into account that $\widehat{S}\in
P\subset \gg_{0}(\s)$, the components of \Gaugetrans\ can be separated as
$$
\eqalignno{
&\left(q^{({\rm can})}(x)\right)_{>0} = \exp\left(\ad \widehat{S}\right)
\biggl(\left(\Lambda +q(x)\right)_{>0}\biggr) - \left(\Lambda\right)_{>0}\>,&
\nameali{Plus}\cr
&\left(q^{({\rm can})}(x)\right)_0 = \exp\left(\ad \widehat{S}\right)
\biggl(\partial_x + \left(\Lambda +q(x)\right)_0\biggr) - \partial_x
-\left(\Lambda\right)_0\>.& \nameali{Zero}\cr}
$$
Obviously $[(\Lambda)_0,P]\subset Q\cap\gg_0(\s)$, and the condition
\Degenplus\
ensures that the map $\ad (\Lambda)_0: P\mapsto Q$ is injective.
Consequently, it is possible to choose the gauge slice by applying the
Drinfel'd-Sokolov prescription to \Zero. Within this choice, the
generator $\widehat{S}$ of the gauge transformation \Gaugetrans\ and the
components of $\left(q^{({\rm can})} (x) \right)_0$ depend only on
$q_0(x)$, while those of $\left(q^{({\rm can})}(x)\right)_{>0}$ depend on both
$q_0(x)$ and $q_{>0}(x)$.

Now, let us analyse in detail the fact that $\ad (\Lambda)_0: P
\mapsto Q_0$ is injective. This means that $\ad (\Lambda)_0:
P^{-j}\mapsto  Q_{0}^{i-j}$ is also injective for any positive integer $j$,
which requires that
$$
{\rm dim\>}\left(\gg_{0}^{-j}\right) \leq {\rm
dim\>}\left(\gg_{0}^{i-j}\right)\>, \quad{\rm for\; any}\quad j\in{\Bbb Z}>0\>.
\nfr{Order}
But ${\rm dim\>} (\gg_{0}^{-j})= {\rm dim\>} (\gg_{0}^{j})$, and it follows
from \Order\ that
$$
{\rm dim\>}\left(\gg_{0}^{-j}\right) = {\rm
dim\>}\left(\gg_{0}^{i-j}\right)\>, \quad{\rm for}\quad 0<j\in {\Bbb
Z}<i\>.
\efr
We conclude that the number of components of $q(x)\in
\gg_0(\s)\cap Q^{>0}$ equals the number of generators of gauge
transformations in $P^{>-i}$, which, under \Degenplus,
ensures the possibility of choosing the gauge slice such that
\Gaugeone\ is satisfied. This completes the proof. $\square$\jump

It is well known that the construction of \W-algebras is based on the
$sl(2,{\Bbb C})$ subalgebras of a finite Lie algebra~[\Ref{EMB}],
while the integrable hierarchies of~[\Ref{GEN1},\Ref{GEN2}] are
constructed in terms of the different Heisenberg subalgebras of a loop
algebra. The form of the reduced current \Reduced\ suggests the kind of
connection between the two constructions that should be investigated.

If $(\Lambda)_0$ does not vanish, it is a nilpotent element of $\gg$. Then
the Jacobson-Morozov theorem  ---theorem 17 of [\Ref{JAC}], pag. 100---
affirms that there exists a unique element $J_-\in \gg_0(\s)\cap
\gg_{-i}(\sw)$ such that
$J_+=(\Lambda)_0$, $J_-$, and $J_0= [J_+,J_-]$ close an $A_1=sl(2,{\Bbb
C})$ subalgebra of $\gg_0(\s)$. $J_0$ and the $H_{\s}$ that defines
\Derivation\ lie on the same Cartan subalgebra of $g$. Their relation
is summarized by the following results.

\jump\noindent
{\bf Lemma 2.} ([\Ref{SORB2}]) {\it Let $S=(J_+,J_0,J_-)$ be the $A_1$
subalgebra of
$\gg_0(\s)$ with $J_+ =(\Lambda)_0$. Then, $H_{\sw} =  J_0
+Y$, and the generator $Y$ commutes with all the generators of S.}
$\square$
\jump

Hence, if $Y$ does not vanish, $J_+=(\Lambda)_0$, $H_{\sw}$,
and $J_-$ generate an $A_1\oplus u(1)$ subalgebra of $\gg_0(\s)$. Under the
adjoint action of this subalgebra, $\gg_0(\s)$, considered as a vector
space, decomposes as the direct sum of a finite number of irreducible
representations of $A_1$
$$
\gg_0(\s) = \bigoplus_{k=1}^{n} D_{j_k}(y_k)\>,
\nfr{Spin}
where $j_k$ and $y_k$ are the spin and the eigenvalue of $Y$ on the
irreducible representation $D_{j_k}(y_k)$. Notice that the
$\sw$-grade of $\Lambda$ is $i$, thus, the normalization of the $A_1$
subalgebra used in lemma~2 is
$$
[J_0, J_{\pm}] = \pm\> i\> J_{\pm}\>,\qquad [J_+,J_-] =J_0\>,
\nfr{Normone}
and the eigenvalues of $J_0$ on $D_{j_k}(y_k)$ are $(-ij_k\> ,\>
-i(j_k-1)\> ,\> \ldots\>, \> +ij_k)$, while those of $H_\sw$ are
$(y_k-ij_k\> ,\> y_k-i(j_k-1)\> ,\> \ldots\>, \> y_k+ij_k)$.

\jump\noindent
{\bf Lemma 3.} ([\Ref{SORB2}]) {\it Under the condition \Degenplus, the
eigenvalues of
$Y$ in the decomposition \Spin\ are bounded according to
$$
\vert y_k \vert\leq\, i\>j_k\>.\qquad \square
\nfr{Bound}
}\sjump

Proposition~2 ensures that the components of $q^{({\rm can})\leq0}(x)$
depend only on $q_0(x)$, but it only specifies the components of the
gauge slice whose $\sw$-grade is $>0$; they  correspond to the gauge
fixing of the transformations generated by $P^{>-i}$. The residual
gauge freedom is generated by the elements of $P^{\leq-i}$, which just
induces the transformation of $q^{\leq0}(x)$ according to \Zero. If
\Degenplus\ is fulfilled, that residual freedom can be fixed using the
Drinfel'd-Sokolov prescription in such a way that
$q^{({\rm can})\leq0}(x)$ is a combination of the lowest weights in the
decomposition \Spin; \ie, of those elements of $Q^{\leq0}$ that cannot
be expressed as $[J_+\> ,\>\ast\>]$, $J_+=(\Lambda)_0$. All this proves the
following.

\jump\noindent
{\bf Theorem 1.} {\it Under the condition \Degenplus, the gauge slice
can be chosen as
$$
q^{({\rm can})}(x)= q^{({\rm can})}_{>0}(x) +
{\cal J}(x)\>,
\nfr{RedPot}
where ${\cal J}(x)$ is a combination of all the lowest weights
of the irreducible representations of $A_1$ into the decomposition
\Spin\ of $\gg_0(\s)$. The components of ${\cal J}(x)$ are polynomial
functionals of $q_0(x)$ and their $x$-derivatives, while those of
$\left(q^{({\rm can})}(x)\right)_{>0}$ depend also on $q_{>0}(x)$.}
$\square$

\sjump\noindent
{\it Remark.} The form of ${\cal J}(x)$ depends only on
$(\Lambda)_0$ because $J_+=(\Lambda)_0$ fixes the $A_1$ subalgebra of
$\gg_0(\s)$ inducing the decomposition \Spin, and lemma~3 ensures that
all the lowest weights of the representations of $A_1$ indeed have not
positive $\sw$-grade. Thus, as far as \Degenplus\ is satisfied, the
form of $\left(q^{({\rm can})}(x)\right)_{0}$ is independent of the
choice of $\sw$. \jump

Let us go back to the form of the energy-momentum tensor. We have
already said that \TTT\ comes from the Sugawara construction applied
to the reduced current \Reduced, improved by a term that depends on the
gradation $\sw$, and that it generates the conformal transformation
\Confone\ on $q_0(x)$. Nevertheless, in general, this energy-momentum tensor
is not gauge invariant. Consider the gauge transformation
$$
q(x)\mapsto \>\widetilde{q}\>(x) =\exp(\ad
S)(\partial_x +\Lambda +q(x)) - \partial_x -\Lambda\>,
$$
then
$$
\eqalignno{
T_\epsilon[\>\widetilde{q}\>] &=
{1\over 2}\int dx\> \epsilon(x)\> \big< \exp(\ad S)\left(\partial_x +
\Lambda
+q(x)\right) - \partial_x\>, \>
\exp(\ad S)\big(\partial_x \cr
& \qquad\qquad \qquad\qquad +\Lambda +q(x)\big) - \partial_x
\big>_{\gg}
\cr  &\qquad+ {1\over i}\int dx\> \epsilon'(x)\>\langle  H_\sw\>, \>
\exp(\ad S)\left(\partial_x + \Lambda +q(x)\right) -
\partial_x\rangle_{\gg}\cr
&= T_\epsilon[q]
- \int dx\> \epsilon(x)\> \langle \exp(-\ad
S)\left(\partial_x\right)-\partial_x
\>,\> \Lambda + q(x)\rangle_{\gg}
\cr &\qquad + {1\over i}\int dx\> \epsilon'(x)\>\langle  \exp(-\ad S)
\left( H_\sw\right) - H_\sw \>, \>\Lambda + q(x)\rangle_{\gg}\cr
&\equiv\>
T_\epsilon[q] \>+ \> \Sigma_\epsilon(S;q)\>,&\nameali{Nongauge} \cr}
$$
where we have used the invariance of the bilinear form of $\gg$. Now,
using \Potential\ and \Gaugegen, it can be checked that
$\Sigma_\epsilon(S;q)$ depends only on the components of $S\in
P^{>-i}$ and on those of $q\in Q^{>0}$. Therefore, $\Sigma_\epsilon
(S;q)$ vanishes when $i=1$ and the energy-momentum tensor
\TTT\ is gauge invariant only in this case.

In contrast, when $i>1$ the energy-momentum tensor \TTT\ is not gauge
invariant. Recall that \TTT\ has been derived from the
condition that it generates the conformal transformation of $q_0(x)$.
As we only need the generator of the conformal transformation of the
gauge invariant functionals of $q_0(x)$, $T_\epsilon[q]$ can be
changed for any other functional ${\>\widehat{T}}_\epsilon[q]$ such
that its functional derivative equals \Tinf\ up to the generator of an
infinitesimal gauge transformation; \ie,
$$
(d_q {\>\widehat{T}}_\epsilon)_0 - (d_q T_\epsilon)_0 \in P\>.
\nfr{Change}
The natural choice is:

\jump\noindent
{\bf Proposition 3.} {\it For any choice of the gauge slice $Q^{({\rm
can})}$, the gauge invariant functional
$$
\Theta_\epsilon[q] = \int dx\> \epsilon(x)\>\Theta(x) =
T_\epsilon[q^{({\rm can})}]\>,
\nfr{Gaugeinv}
generates the conformal transformation of the gauge invariant functionals of
$q_0(x)$.}

\sjump\noindent
{\it Proof.} To start with, let us recall that the components of $q^{({\rm
can})}(x)$ are understood as gauge invariant functionals of $q(x)$, and that
we fix the gauge following the Drinfel'd-Sokolov prescription. The
relation between the components of $q^{({\rm can})}(x)$ whose $\sw$-grade is
$j$ and
$q(x)$ is
$$
\eqalign{
q^{({\rm can})j}(x) & = \biggl(\exp (\ad {\>\widehat{S}\>})
\left(\partial_x + \Lambda +q(x)\right) - \partial_x -
\Lambda\biggr)^j \cr &= q^j(x) - [\Lambda, {\>\widehat{S}\>}^{\>j-i}(x)] +
\Delta({\>\widehat{S}\>}^{\>>j-i}; q^{>j})\>.\cr}
\nfr{Gaugefixing}
The components of ${\>\widehat{S}\>}$ are functionals of $q(x)$ fixed by the
condition that $q^{({\rm can})}(x)\in Q^{({\rm can})}$, with $Q^{({\rm can})}$
some complementary vector space of $[\Lambda,P]$ in $Q$;
so, ${\>\widehat{S}\>}^{-j}(x)$ depends only on the components of $q(x)$ whose
$\sw$-grade is $\geq i-j$. Moreover, \Gaugefixing\ also shows that
$q^{({\rm can})j}(x)$ is a functional only of the components of $q(x)$
with $\sw$-grade $\geq j$. Now, let us consider \Nongauge\ with
$S={\>\widehat{S}\>}$; since $\Sigma_\epsilon({\>\widehat{S}\>};q)$ depends
only on the components of $q(x)$ whose $\sw$-grade is positive, we
conclude that $(d_q
\Sigma_\epsilon ({\>\widehat{S}\>};q))_0 \in P$, which, using \Change, proves
the proposition. $\square$\jump

Finally, with the choice of the gauge slice of theorem~1, the
gauge invariant energy-momentum tensor \Gaugeinv\ becomes
$$
\eqalign{
\Theta(x) & = {1\over 2}\langle \Lambda + q^{({\rm can})}(x)\>, \>
\Lambda + q^{({\rm can})}(x)\rangle_{\gg}
\,
- \, {1\over i} \partial_x\langle  H_\sw\>, \>  \Lambda +
q^{({\rm can})}(x)\rangle_{\gg}\cr
&= \langle \left(\Lambda \right)_0\>, \> {\cal J}^{-i}(x)\rangle_{\gg}
\, +\, {1\over 2}\langle {\cal J}^{0}(x)\>, \> {\cal
J}^{0}(x)\rangle_{\gg}\, -
\, {1\over i} \partial_x\langle  H_\sw\>, \>  {\cal J}^{0}(x)
\rangle_{\gg}\> ,\cr}
\nfr{Tgauge}
which shows that $\Theta(x)$ is a gauge invariant functional of
$q_0(x)$.

\chapter{{\bf The second Poisson bracket and the \W-algebras.}}

Theorem~1 shows that, under condition \Degenplus, the
components of ${\cal J}(x)$ provide a basis for the gauge invariant
functionals of  the components of $q(x)$ whose $\s$-grade is zero.
Obviously, the second Poisson bracket of two of these functionals is
another gauge invariant functional of $q_0(x)$ and it makes
sense to consider the restriction of the second Poisson bracket algebra
to these functionals. Recall that, in the previous section, the
components of $q^{({\rm can})}(x)$ have to be understood as functionals
of $q(x)$, hence it would be more precise to write eq.\RedPot\ as
$$
q^{({\rm can})}\biggl(q(x)\biggr)= q^{({\rm
can})}_{>0}\biggl(q_0(x),q_{>0}(x)\biggr) +  {\cal J}\biggl(q_0(x)
\biggr)\>,
\nfr{RedPotf}
and it follows from the gauge fixing prescription that $q^{({\rm
can})}\left(q(x)\right) = q(x)$ if $q(x)\in Q^{({\rm can})}$; in
particular, ${\cal J}\left(q_0(x)\right)= q_0(x)$ if $q_0(x) \in
Q^{({\rm can})}_0$.

In the previous section, we have proved that the
restriction of the second Poisson bracket algebra to the gauge invariant
functionals of $q_0(x)$ is just the Hamiltonian reduction of the
untwisted affinization of $\gg_0(\s)$ defined by the bracket \KM, and
corresponding to
$$
J(x) = \left(\Lambda\right)_0 + {\cal J}(x)\>,
\nfr{Redy}
which is one of the \W-algebras considered
in~[\Ref{EMB},\Ref{HAM2},\Ref{SORB1}]. In particular, it is the
\W-algebra associated to the $A_1$ subalgebra of $\gg_0(\s)$ whose
nilpotent $J_+$ element is $\left(\Lambda\right)_0$.

Notice that the relevant conformal structure in terms of which all the
generators of the $\cal W$-algebra are primary fields is defined by the
semisimple element of the $A_1$ subalgebra $J_0$, and the appropriated
energy-momentum tensor is
$$
{\cal T}(x) = \langle \left(\Lambda\right)_0\>, \> {\cal
J}^{-i}(x)\rangle_{\gg}\, +\, {1\over 2}\langle {\cal J}^{0}(x)\>, \>
{\cal J}^{0}(x)\rangle_{\gg}
\>,
\nfr{Tfinal}
such that~[\Ref{HAM2}]
$$
\big\{ {\cal T}_\epsilon[q], {\cal J}(x) \big\}= \epsilon(x)  {\cal
J}'(x) +
\epsilon'(x) \left( {\cal J}(x) - {1\over i}\left[J_0\>, \>{\cal
J}(x)\right]\right) - {1\over i} \epsilon'''(x) J_-\>,
\nfr{Primary}
where we use the normalization \Normone, and ${\cal T}_\epsilon[q]=\int
dx\> \epsilon(x)\> {\cal T}(x)$. ${\cal T}(x)$ closes a
classical Virasoro algebra whose central extension is $12\langle
J_+,J_-\rangle/i=12\langle J_0,J_0\rangle/i^2$. As the elements of ${\cal
J}(x)$ are the lowest weights in \Spin, $\langle  J_0, {\cal J}(x)
\rangle_{\gg}=0$ and ${\cal T}(x)$ equals $\Theta(x)$ only if $H_{\sw}$ is
proportional to $J_0$; otherwise,
$$
{\cal T}(x) - \Theta(x) = {1\over i} \partial_x \langle H_{\sw} -
J_0\>, \> {\cal J}^0(x)\rangle_{\gg}\>.
\efr
$\Theta(x)$, given by \Tgauge, defines the
conformal structure corresponding to the transformation \Conformal\ that
generalizes the  scaling transformation
\Scaling\ under which the integrable hierarchy is homogeneous. Hence,
in general, the conformal structure of the
\W-algebra is twisted with respect to the conformal structure defining
the scaling properties of the integrable hierarchy.

In relation to this, it is convenient to consider again the remark
below theorem~1. It indicates that, as long as \Degenplus\ keeps up, the
second Poisson bracket algebra does not depend on the choice of the
gradation $\sw$. We have anticipated in Section~2 that, for a given
Heisenberg subalgebra $\Heis$, the choice of $\sw$ is not unique, a
question that has not been mentioned at all in~[\Ref{GEN1},\Ref{GEN2}].
The origin of this ambiguity can be easily understood by noticing that,
for a given choice of
$\sw$, the elements of
$\Heis$ are also graded by any other gradation $\s'$ such that
$$
{H_{\sw} \over N_{\sw}} - {H_{\s'}\over N_{\s'}} \in \Heis\>.
\nfr{Ambiguity}
In fact, it can be proven that these are the only ambiguities in
the choice of $\sw$, and that they are obviously linked to the existence of
elements of $\Heis$ whose $\sw$-grade is zero. Moreover, notice that
\Ambiguity\ ensures that the grade of the elements of $\Heis$ is
uniquely defined up to multiplication by a global rational factor, in
agreement with lemma~1. This shows, in particular, that the scaling
dimensions of the times $t_{b_j}$ in \Scaling\ are also uniquely defined. Let
us consider the Heisenberg subalgebras of $\gg$ as the affinization in $z$ of
a fixed Cartan subalgebra of $g$~[\Ref{KP}]. The different Heisenberg
subalgebras arise in relation to the different finite order (inner)
automorphisms of $g$, which are classified by the conjugacy classes of
the Weyl group of $g$~[\Ref{KP}]. The connection between a Weyl group
element, say $w$ up to conjugacy, and the associated Heisenberg
subalgebra $\Heis$ is that  there is a lift of $w$ onto $\gg$, denoted
$\widehat{w}$,  which acts on $\Heis$ as
$$
\widehat{w}(b_j) \equiv w(b_j) =\exp \left({2\pi i\over N}\>j \right)
b_j\>,
\nfr{Lift}
with $N$ being the order of $w$, $w^N =1$. The different choices of $\sw$
correspond precisely to the different  choices of
$\widehat{w}$, which, in the ``shifted picture'', can be expressed
as~[\Ref{KACB}]
$$
\widehat{w} = \exp \left({2\pi i\over N_{\sw}}\> \ad H_\sw\right)\>.
\nfr{Shift}
The choice of $\sw$ is unique for the principal Heisenberg
subalgebra, which is related to the conjugacy class of the Coxeter
element of the Weyl group of
$g$. In contrast, the choice of $\sw$ for the homogeneous Heisenberg
subalgebra, for which $w$ is the identity, is completely arbitrary.
In~[\Ref{KP}], a canonical choice of $\sw$ is determined by
$$
\langle\> H_\sw\,, \, b_j\>\rangle = 0\>, \quad {\rm for} \quad  {\rm
any} \quad b_j\in\Heis\>;
\nfr{Kaccond}
his condition is non-trivial only for $b_j\in \Heis\cap
\gg_0(\sw) \subset \Heis\cap \gg_0(\s)$. Let us assume that $H_\sw\in
{\rm Im}(\ad(\Lambda)_0)$,
\ie, that there  exists some $M\in\gg_0(\s)\cap\gg_{-i}(\sw)$ such that
$H_\sw= [(\Lambda)_0,M]$. Then, the invariance of the
bilinear form of $\gg$ implies that
$\langle\> H_\sw\,, \, b_j\>\rangle= \langle\> [(\Lambda)_0,M]\,, \,
b_j\>\rangle$ vanishes and \Kaccond\ can be viewed as a necessary
condition to ensure that $H_{\sw}$ is proportional to $J_0$.\sjump

All these results fix the connection
between the algebras defined by the second Poisson bracket of the
integrable hierarchies of~[\Ref{GEN1},\Ref{GEN2}] and the \W-algebras
of~[\Ref{EMB},\Ref{HAM2},\Ref{SORB1}].

\jump\noindent
{\bf Theorem 2.} {\it Let us consider the integrable hierarchy
associated to  the data
$$
\{\gg,\Heis,s_w,\s,\Lambda\}\>,\;\>
such\;\> that\;\> {\rm Ker}(\ad(\Lambda)_0) \cap P
=\emptyset\>.
$$ (i)~If $(\Lambda)_0\not=0$, the restriction of
the second Poisson bracket algebra to the gauge invariant functionals
of $q_0(x)$ is the \W-algebra associated to the $A_1$ subalgebra of
$\gg_0(\s)$ specified by the nilpotent element $J_+=(\Lambda)_0$.
(ii)~Otherwise, if $(\Lambda)_0=0$, such restriction is just the
Kac-Moody algebra  corresponding to the untwisted affinization of
$\gg_0(\s)$.}
$\square$\jump

Together with this theorem, let us point out, first, that theorem~2 does not
specify the Poisson  brackets involving the components of $ q^{({\rm
can})}_{>0}(x) $; in any case, proposition~1 shows that some of these
components are centres of the Poisson bracket algebra. Second,
in general, the gradation $\sw$ cannot be chosen such that
$H_{\sw}$ is proportional to $J_0$; therefore, the conformal
transformations generated by
${\cal T}(x)$ will be different from the transformation induced by
\Confone\ and \Conformal. Anyway, it follows from lemma~2 that
the condition \Degenplus\ is always satisfied in those particular cases when
$H_{\sw}$ is proportional to $J_0$. So, it is interesting to consider
the following straightforward result.

\jump\noindent
{\bf Lemma 4.} {\it Let us suppose that $H_{\sw}$ is proportional to
$J_0$. Then, either $i=1$ or $i=2$, and the value
$i=1$ means that the corresponding embedding of $A_1$ into $g$ is
integral.}

\sjump\noindent
{\it Proof.} Let us look at the decomposition \Spin. According to
lemma~3, and with the normalization used there, $H_{\sw}$ being
proportional to
$J_0$ means that, actually, $H_{\sw}=J_0$, or, equivalently, that $Y=0$.
Then, the gradation induced by $\sw$ is
$$
\gg= \bigoplus_{k\in I} \gg_k(\sw)\>,
$$
where
$$
I=\bigcup_{k=1}^{n}\{ -ij_k,-i(j_k -1),\ldots, i(j_k -1),
ij_k\} \>+ \>{\Bbb Z}\> N_{\sw}\>,
$$
which means that all the grades are proportional to $i$. Since $\sw$ is an
integer gradation of $\gg$, and the non-zero grades are relatively
prime, there are only two possibilities. The first one is that all the
representations in \Spin\ have integer spin; in this case
$i=1$. The second possibility is that there are representations whose
spin is half-integer, then $i=2$. $\square$\jump

Theorem~2 generalizes the well known relationship between the
Gel'fand-Dickey algebras and the \W-algebras~[\Ref{FL},\Ref{BAK}] that
we have summarized in the introduction. The logical question is if all
the \W-algebras that can be constructed out of the inequivalent $A_1$
subalgebras can be recovered in this way. This requires that, for
any $A_1$ subalgebra of $\gg_0(\s)$, $J_+$ can be expressed as the zero
$\s$-grade projection of a $\sw$-graded element of some Heisenberg
subalgebra of $\gg$. Hence, it is necessary to investigate in detail
the relation between the embeddings of $A_1$ into a Lie algebra $g$ and
the conjugacy classes of the Weyl group of $g$ that classify the
inequivalent Heisenberg subalgebras of $g^{(1)}$.
A formal relation of this kind has been realized by
mathematicians~[\Ref{CARELK}] after the classification of the
conjugacy classes of the Weyl group was completed, and, more recently,
this topic has attracted the interest of physicists in connection
with the structure of integrable two-dimensional
field theories~[\Ref{UNDER},\Ref{FRENK}].

Before addressing this question in the next section, let us settle
the two following results that will be useful when $\s$ is the homogeneous
gradation.

\jump\noindent
{\bf Lemma 5.} {\it The components of $\gg$ whose homogeneous grade is
$\geq m$  have $\sw$-grade $\geq ({\sw})_{0} + (m-1) N_\sw$.}

\sjump\noindent
{\it Proof.} Notice that the $\sw$-grade of any element of $g$ ($\equiv
g\otimes z^0$) is $\geq$ than the grade of the lowest root's step
operator, \ie, $\geq$ than
$$
-\sum_{j=1}^{r} k_j ({\sw})_j = ({\sw})_{0} - N_\sw\>,
\efr
where we have used that the Kac label $k_0=1$ for the untwisted
Kac-Moody algebras. The result follows just by considering the
$\sw$-grade of the elements $a\otimes z^n\in \gg$ with $n\geq m$.
$\square$

\jump\noindent
{\bf Lemma 6.} {\it The minimal positive $\sw$-grade of the elements of
$\Heis$ is $\geq (\sw)_0$. Moreover, if $\sw$ is $\succeq$ than the
homogeneous gradation, then $(\sw)_0\geq1$.}

\sjump\noindent
{\it Proof.} To define the gradations, we have chosen a
particular Cartan subalgebra $H$ of $g$ in such a way that the
elements of $H$ have zero grade with respect to any gradation (see
\Derivation). Besides, if $B_{\pm}$ are the maximal positive and negative
nilpotent subalgebras of $g$ with respect to $H$,
\ie, $g=B_+\oplus H\oplus B_-$, the elements of $B_+$ and $B_-$ have,
respectively, grade
$\geq0$ and
$\leq0$ with respect to whatever gradation. Recall now
that  the elements of $\Heis$ are semisimple while those of $B_\pm$ are
nilpotent. Hence, any element of $\Heis$ with positive $\sw$-grade has
to be of the form $z^n\otimes\alpha_+\>  +\> z^{n+1}\otimes\alpha_{-}$,
with $n\geq0$, and $\alpha_\pm\in B_\pm\not=0$. Therefore, the first
part  of the lemma follows by using lemma~5 with $m=1$, and if $\sw$ is
$\succeq$ than the homogeneous gradation, then
$(\sw)_0\not=0$, \ie, $(\sw)_0\geq1$. $\square$\jump

These two lemmas lead to

\jump\noindent
{\bf Lemma 7.} {\it When $\s$ is the homogeneous gradation,
$(\Lambda)_0\not=0$ if and only if
$$
(\sw)_0 \leq i < N_\sw\>.
\nfr{GenKdV}
Moreover, if $i=(\sw)_0$, the $\s$-grade of all the components of
the potential $q(x)$ is zero, \ie, $q(x) \in \gg_0(\s)$, and the
two conditions \Degen\ and \Degenplus\ are equivalent.}
$\square$\jump

As a remark, let us mention that, in general, not all Heisenberg
subalgebras have generators whose $\sw$-grade equals the lower
bound $i=(\sw)_0$.

\chapter{{\bf Embeddings of $A_1$ and conjugacy classes of the Weyl
group.}}

In this section, we restrict ourselves to the $A_n$ algebras. Similar
results for the other classical Lie algebras will be presented
elsewhere.

\section{{\bf The $A_1$ subalgebras of $A_n$.}}

The embeddings of $A_1$ into a semisimple Lie algebra have been
classified originally by E.B.~Dynkin~[\Ref{DYN}], a classification that
has been made explicit for those algebras of rank up to $6$
in~[\Ref{LG}] ---see also~[\Ref{SORB1},\Ref{RAG}] for recent reviews of
this classification inspired by the construction of $\cal W$-algebras
and non-abelian Toda models. Here, we normalize the $A_1$
subalgebras as
$$
[J_0, J_{\pm}] = \pm\> 2\> J_{\pm}\>,\qquad [J_+,J_-] =J_0\>.
\nfr{Normtwo}
The different embeddings of $A_1$ into the Lie algebra $g$ correspond
to the different injective homomorphisms $\imath: A_1 \mapsto g$, and,
in practice, we shall identify $A_1$ with its image under $\imath$. An
embedding of $A_1$ is specified by its defining vector
${\bf f} =(f_1,\ldots,f_r)$, which provides the decomposition of $J_0$
with respect to a basis of the Cartan subalgebra
$\{H_1,\ldots,H_r\}$ of $g$,
$$
J_0 = \sum_{j=1}^{r} f_j H_j\>;
\nfr{Defining}
obviously, $r$ is the rank of the Lie algebra $g$.
It can be proved that two embeddings of $A_1$ are conjugated if and
only if their $J_0$'s are conjugated under some automorphism of
$g$~[\Ref{DYN}]. Even more, through conjugation with the Weyl group of
$g$, the basis of simple roots $\{\alpha_1,\ldots,\alpha_r\}$ can be
chosen such that the quantities $c_j=\alpha_i(J_0)\geq0$; in
fact, with the normalization \Normtwo, these numbers can only take the values
$0,\;1$, or
$2$. Considering the decomposition of the Lie algebra $g$,
as a vector space, under the adjoint action of $\imath(A_1)$, the
value of $c_j=2,\;1$, or $0$ indicates that the step operator
$E_{\alpha_j}$ belongs to a representation whose spin is integer,
half-integer, or zero, respectively. The numbers $c_1,\ldots,c_r$ are
associated to the nodes of the Dynkin diagram of $g$, and this set of
numbers is called the ``characteristic'' of the embedding.

A particularly important example is provided by the ``principal
embedding'' defined as~[\Ref{KOST}]
$$
J_{+}^{(\rm p)} = \sum_{j=1}^{r} \lambda_{j} E_{\alpha_j}\>, \qquad
J_{0}^{(\rm p)} = \sum_{j=1}^{r} {4\>\omega_j\over \alpha_{j}^{2}}
\cdot H\>,
\nfr{Princemb}
where the $\omega_j$'s are the fundamental weights of $g$ defined as
$\alpha_j\cdot \omega_k = (\alpha_{j}^{2}/2)\delta_{jk}$. The
$\lambda_j$'s are non-vanishing arbitrary constants, usually taken to
be $1$, but notice that $J_0$ is completely independent of them. The
``characteristic'' of the principal embedding is
$\{2,2,\ldots,2\}$.

For $A_n$, it is convenient to represent the Cartan subalgebra using an
$(n+1)$-dimensional Euclidean space as follows~[\Ref{SORB1}]. Let
$\{e_1,\ldots,e_{n+1}\}$ be a basis of this space that is orthonormal
with respect to the Cartan-Killing form. In this basis,the roots
of $A_n$ are $(e_i - e_j)$, where $i,j=1,\ldots,n+1$, and a suitable choice of
the  simple roots is $\{\alpha_j = e_j-e_{j+1},
\;j=1,\ldots,n\}$. Accordingly, the basis of the Cartan subalgebra
can be chosen in terms of the set
$\{H_1,\ldots,H_{n+1}\}$ such that
$$
[H_i, E_{(e_j-e_k)}] = (\delta_{i,j} -\delta_{i,k}) E_{(e_j-e_k)}\>,
\qquad i,j,k=1,\ldots,n+1\>,
\nfr{Basis}
and the elements of the Cartan subalgebra are
$\sum_{j=1}^{n+1} p_j H_j$ with $\sum_{j=1}^{n+1} p_j =0$. Having made these
choices, the defining vector of the principal embedding of $A_1$ into
$A_n$ is
$$
{\bf f}^{\>\rm p}= \left( n,n-2,\ldots, 2-n,-n\right)\>.
\nfr{Embdef}

Dynkin's result is that all the non-conjugated embeddings of
$A_1$ into $A_n$ can be described as the principal $A_1$ embedding of a
regular subalgebra\note{Consider the root space decomposition of a Lie
algebra $g$ with respect to a Cartan subalgebra $H$:
$g= H\> \oplus_{\alpha \in \Delta} g_\alpha$,
where $\Delta$ is the root system of $g$. Then, a subalgebra
$g'\subset g$ is called regular if it can be seen, up to conjugation, as
$g'= H' \oplus_{\alpha \in \Delta'} g_{\alpha}$,
with $\Delta'\subset \Delta$ and $H' \subset
H $.} of
$A_n$.  Dynkin has also classified the regular subalgebras of semisimple Lie
algebras~[\Ref{DYN}]. In the case of $A_n$, they are in
one-to-one relation with the partitions of
$n+1$, and the regular subalgebra of $A_n$ associated to
$$
n+1 = \sum_{j=1}^{k} n_j \>,\quad n_k\geq n_{k-1}\geq \cdots \geq
n_1\>,
\nfr{Partone}
where $k\geq1$ and $n_j$ are positive integer numbers, is
$$
\bigoplus_{j=1\atop n_j\not=1}^{k}\> A_{n_j-1} \subset A_n\>.
\nfr{Subalg}
Then, the corresponding embedding of $A_1$ into $A_n$ is defined by
$$
J^{(n_k,\ldots,n_1)}_{\pm} = \sum_{j=1\atop n_j\not=1}^{k} \>
\left(J_{\pm}^{({\rm p})}\right)_{ A_{n_j-1}}\>, \qquad
J^{(n_k,\ldots,n_1)}_{0} = \sum_{j=1\atop n_j\not=1}^{k}
\left(J_{0}^{({\rm p})}\right)_{ A_{n_j-1}}\>.
\nfr{Diagemb}
The defining vector of this embedding is just the union of
the defining vectors of the principal embeddings of each $A_{n_j-1}$
factor in \Subalg\ plus the required number of zeroes to complete the
$n+1$ components of {\bf f} ---the number of zeroes equals the number
of $1$'s in \Partone. The components of the resulting {\bf f}
can be arranged such that $f_{n+1} \leq f_n
\leq \cdots \leq f_1$, which corresponds to the change of the basis of
simple roots leading to the ``characteristic'' of the
embedding~[\Ref{SORB1}]. Finally, after the components of $\bf f$ have
been ordered, the components of the ``characteristic'' are $c_j=f_j -
f_{j+1}$, $j=1,\ldots,n$.

For instance, the defining vector of the embedding associated to the
partition $4=2+1+1$, \ie, to the regular subalgebra $A_1\subset A_3$,
is $(1,0,0,-1)$. Another example, the  defining vector corresponding to
$7=4+2+1$, \ie, to $A_3\oplus A_1\subset A_6$, is $(3,1,1,0,-1,-1,-3)$,
as follows from  the union of $(3,1,-1,-3)$, $(1,-1)$, and $(0)$. In
these two examples, the ``characteristics'' are $(1,0,1)$
and $(2,0,1,1,0,2)$, respectively.

Notice that, for any embedding of $A_1$ into $A_n$, the components of
{\bf f} satisfy the symmetry relation $f_j=  -f_{n+2-j}$, which implies
that the components of the ``characteristic'' satisfy $c_j=c_{n+1-j}$.

\section{{\bf The conjugacy classes of the Weyl group of $A_n$ and the
Heisenberg subalgebras of $A_{n}^{(1)}$.}}

It has been shown by V.G.~Kac and D.H.~Peterson that the inequivalent
Heisenberg subalgebras of the affine untwisted Kac-Moody algebra
$g^{(1)}$ are in one-to-one relation with the conjugacy classes of the
Weyl group of $g$~[\Ref{KP}] ---we have briefly summarized the
connection between the elements of the Weyl group of $g$ and the
Heisenberg subalgebras of $g^{(1)}$ in \Lift\ and
\Shift. The complete classification of the conjugacy classes of the
Weyl group, including classical and exceptional Lie algebras, has been
developed by R.W.~Carter~[\Ref{CART}]. In connection
with the construction of the twisted vertex operator representations
of Kac-Moody algebras, more recently, very detailed descriptions of the
conjugacy classes of the Weyl group and the Heisenberg subalgebras of the
untwisted affinizations of the classical Lie algebras have been
presented in~[\Ref{KROOD1}, \Ref{KROOD2}] (see also~[\Ref{FHM},\Ref{DFEH}]).

The Weyl group of $A_n$ is isomorphic to $S_{n+1}$, the group of
permutations of $n+1$ elements. Therefore, it can be represented as a
set of linear transformations acting on an $(n+1)$-dimensional
Euclidean space with an orthonormal basis $\{e_1,\ldots,e_{n+1}\}$
---the same we have used to describe the Cartan subalgebra of
$A_n$. The elements of the Weyl group of $A_n$ are permutations of the
elements of this basis; it can be shown that the conjugacy
classes of the Weyl group of $A_n$ correspond to permutations without
specifying the particular sets of $e_j$'s on which they act. The
conjugacy classes of the Weyl group of $A_n$ are too in one-to-one
relation with the partitions of $n+1$. The conjugacy class
corresponding to the partition \Partone\ is usually denoted as
$[n_1,n_2,\ldots,n_k]$, where the order of the $n_j$'s is irrelevant.
If the elements of the basis are divided into $k$  subsets with
$n_1,\,n_2,\ldots,\,n_k$ elements, respectively, the conjugacy class
consists in a cyclic permutation in each of the subsets, \ie,
\vskip 0.05cm
$$
\eqalign{
& e_1 \rightarrow e_2\rightarrow\cdots \rightarrow e_{n_1}\rightarrow
e_1 \cr  & e_{n_1+1} \rightarrow e_{n_1+2}\rightarrow\cdots
\rightarrow e_{n_1+n_2}\rightarrow e_{n_1+1}\cr
& \cdots\cdots\cdots\cdots\cdots \cr
& e_{n_1+\ldots+n_{k-1}+1} \rightarrow
e_{n_1+\ldots+n_{k-1}+2}\rightarrow\cdots \rightarrow e_{n+1}
\rightarrow e_{n_1+\ldots+n_{k-1}+1} \>.\cr}
\nfr{Permut}
\vskip 0.1cm

A particularly important element of the Weyl group is the Coxeter
one, and the conjugacy class of the Coxeter element of $A_n$ is
just $[n+1]$. It  can be shown that the eigenvalues of any
element of $[n+1]$ are
$\{\omega^j, \;j=1,\ldots,n\}$, with $\omega$ being an $(n+1)$-root of
$1$, $\omega^{n+1}=1$.

In the general case, the conjugacy class
$[n_1,n_2,\ldots,n_k]$ is simply the conjugacy class of the Coxeter
element of the regular subalgebra \Subalg\ associated to the
partition \Partone. By the Coxeter element of a semisimple Lie algebra we
mean just the composition of the Coxeter elements of each simple ideal.
Given a regular subalgebra of a simple Lie algebra $g$, the Cartan subalgebra
of $g$ can be constructed as the union of the Cartan
subalgebras of each simple ideal of the subalgebra plus some additional
elements to match the rank of $g$; obviously, the Coxeter element of the
regular subalgebra leaves these additional elements invariant. In the case of
\Subalg, the number of invariant directions is precisely $k-1$, as
follows from the difference between the rank of the right-hand-side
($n$) and of the left-hand-side ($\sum_{j=1}^{k}(n_j-1) = n+1-k$).

As summarized by \Lift\ and \Shift, to construct the Heisenberg subalgebra of
$\gg$ associated to a conjugacy class of the Weyl group of $g$, one needs
a lift of a representative element of the class onto
$\gg$; the lift of the Coxeter element is uniquely defined. For $A_n$,
it can be expressed as
$$
\widehat{w}_{\rm c} = \exp \left( {2\pi i\over 2(n+1)}\>  \ad
J_{0}^{({\rm p})}
\right) \>,
\nfr{LiftCox}
where $J_{0}^{({\rm p})}$ has been given in \Princemb. The Heisenberg
subalgebra of $\gg$ associated to the conjugacy class of the Coxeter element
is called the principal Heisenberg subalgebra, and it can be defined as the
centralizer in $\gg$ of the affine ``cyclic element'' of
B.~Kostant~[\Ref{KOST}],
$$
b^{(K)}= \sum_{j=1}^{r} \lambda_j\>1\otimes E_{\alpha_j} + \lambda_0\> z
\otimes
E_{-\psi}\>,
\nfr{Konstant}
where the $\lambda_j$'s are arbitrary non-vanishing coefficients, usually
taken to be $1$. $\psi=\sum_{j=1}^{r} k_j \alpha_j$ is the highest root of
$g$, and $k_0=1$, $k_1,\ldots$, and $k_r$ are the Kac labels of $\gg$.
Regarding $A_n$ and using the fundamental representation of
$sl(n+1,{\Bbb C})$, the affine cyclic element is the well known $(n+1)\times
(n+1)$  matrix
$$
\left( b^{(K)}\right)_{A_n}=\pmatrix{0&1& & & \cr
          &0&1& & \cr
          & &\ddots&\ddots& \cr
          & & &0&1\cr
         z& & & &0\cr}\>,
\nfr{Lambdotra}
and the elements of the principal Heisenberg subalgebra are
$$
b^{({\rm p})}_{j+(n+1)\,m}=z^m\> \pmatrix{0& {\Bbb I}_{n+1-j}\cr
               z\>{\Bbb I}_{j}& 0}\>;\qquad
m\in{\Bbb Z}\>, \quad j=1,\ldots,n\>,
\nfr{PrinHeis}
where $j+(n+1)m$ is the principal grade of this element, ${\Bbb I}_j$
is the
$j\times j$ identity matrix, and $b^{({\rm p})}_1 = b^{(K)}$.
Obviously,  the associated gradation is also uniquely defined; it is
the principal gradation $\s_{\rm p}=(1,1,\ldots,1)$ that is induced by
$H_{\s_{\rm p}}= J_{0}^{({\rm p})}/2$.

We have already indicated that a conjugacy class $[w]$ of
the  Weyl group of $A_n$ is the conjugacy class of the Coxeter element
of some of its regular subalgebras, say $\widetilde{g}$. Therefore, a possible
lift $\widehat{w}$ is obtained by composing the lifts of the Coxeter
elements of the simple ideals of $\widetilde{g}$. For instance, for the
conjugacy class $[w]=[n_1,n_2,\ldots,n_k]$ the lift would be
$$
\widehat{w} = \prod_{j=1 \atop n_j\not=1}^{k}
\exp \left( {2\pi i\over 2n_j}\> \ad
\left(J_{0}^{({\rm p})}\right)_{A_{n_j-1}}\right)\equiv
\exp\left({2\pi i\over N_{\sw}} \> \ad H_{\sw}\right)\>,
\nfr{Liftgen}
where we have used the same notation as in \Shift. The order of
$\widehat{w}$ is the following~[\Ref{KROOD1}]: let
$N_{\sw}'$ be the least common multiple of $(n_1,n_2,\ldots,n_k)$, then
$$
N_{\sw} = \cases{N_{\sw}'\>,&if $N_{\sw}'\left({1\over n_i} + {1\over
n_j}\right) \in 2{\Bbb Z}\>$ for any $i,j=1,\ldots,k\>$;\cr
2N_{\sw}'\>,&otherwise.\cr}
\nfr{Double}
Moreover,
$$
H_{\sw} = \sum_{j=1\atop n_j\not=1}^{k} {N_{\sw} \over 2n_j}\>
\left(J_{0}^{({\rm p})}\right)_{ A_{n_j-1}}\>.
\nfr{Gradator}
The decomposition of $H_{\sw}$ with respect to the basis \Basis\ of the
Cartan subalgebra would be
$$
H_{\sw}= \sum_{j=1}^{n+1} \gamma_j H_j\>,\qquad \sum_{j=1}^{n+1}
\gamma_j=0\>;
\nfr{Pregrad}
the $\gamma_j$'s can be arranged such that $\gamma_{n+1}\leq
\gamma_n\leq \cdots\leq \gamma_1$, which corresponds to a change of the basis
of simple roots similar to the one used to order the components of the
defining vectors \Defining\ of the embeddings of $A_1$ into $A_n$.
Once ordered, they have the symmetry property $\gamma_j =
-\gamma_{n+2-j}$, and it follows from \Embdef\ that
$\gamma_1=-\gamma_{n+1}= N_{\sw}(n_k -1)/2n_k$, where $n_k$ is the
maximum of $n_1,\ldots,n_k$.
$H_\sw$ induces an integer gradation of $A_{n}^{(1)}$, $\sw$, whose
components are
$$
\eqalign{
({\sw})_j &= \gamma_j - \gamma_{j+1}= ({\sw})_{n+1-j}\>, \qquad j=1,
\ldots,n\cr ({\sw})_0 &=  N_\sw - \sum_{j=1}^{n} ({\sw})_j = N_\sw -
2\gamma_1 = {N_\sw\over n_k}\>.\cr}
\nfr{Gradation}
For example, the gradation of $A_{3}^{(1)}$ associated to the
conjugacy class $[1,1,2]$ is
$(2,1,0,1)$, while the gradation of $A_{6}^{(1)}$ associated to
$[1,2,4]$ is
$(2,1,1,1,1,1,1)$; it is useful to compare these gradations with the
``characteristics'' of the embeddings associated to the same partitions.

The Heisenberg subalgebra associated to the conjugacy class
$[w]=[n_1,n_2,\ldots,n_k]$ is obtained as follows. Consider the Cartan
subalgebra defined by joining the Cartan subalgebras of
each simple factor in \Subalg\ plus the required $k-1$ additional
generators. The corresponding Heisenberg subalgebra is just the
affinization of this  Cartan subalgebra according to the automorphism
\Liftgen. Then, it will be generated by the linear combinations of the
principal Heisenberg subalgebras associated to each of the simple
factors in \Subalg, embedded in $A_n$, plus the affinization of the
$k-1$ additional generators~(see also~[\Ref{FHM},\Ref{KROOD1}]). As a
practical example, we present all the inequivalent
Heisenberg subalgebras of $A_{2}^{(1)}$ and $A_{3}^{(1)}$ in the
appendix. By construction, the Heisenberg subalgebra has a basis whose
generators are graded by the gradation $\sw$ defined in \Gradation, and
their grades follow from those of the principal Heisenberg
subalgebras of each simple ideal in \Subalg. They are\note{Notice that
this essentially agrees with the results of ref.[\Ref{UNDER}] except that the
possibility of $N_{\sw}$ being twice the least common multiple of
$\{n_1,\ldots,n_k\}$ has not been considered there. Then, the condition to
have a grade~1 element in $\Heis$ is that $N_{\sw}=n_k$, the maximun of
$n_1,\ldots,n_k$. This requires not only that $n_k/n_j \in {\Bbb Z}^+$ for
any $j=1,\ldots,k$, as already realized in~[\Ref{UNDER}], but also that
either all the $(n_k/n_j)$'s are even or that all of them are odd.}
$$
\underbrace{0,\>\ldots,\>0}_{k-1\rm \; times}, \bigcup_{j=1 \atop
n_j\not=1}^{k}
\bigg\{{N_{\sw}\over n_j},{2N_{\sw}\over n_j}\ldots,{(n_j-1)N_{\sw}
\over n_j}\bigg\}{\rm
\;\; mod\>}(N_\sw)\>,
\nfr{Grades}
which are obviously integers because of \Double.
Notice that, also by construction, the $H_{\sw}$ of the lift \Liftgen\
satisfies the condition \Kaccond.

It follows from \Grades\ that the lowest positive grade of the
generators of  the Heisenberg subalgebra of $A_{n}^{(1)}$ associated to
the conjugacy class
$[w]=[n_1,n_2,\ldots,n_k]$ is $N_{\sw}/n_k$, which,
taking into account \Gradation, agrees with the result of lemma~6.
Moreover, it also follows from
\Princemb, \Diagemb, and \Konstant\ that the generator $J_+$ of the
embedding of
$A_1$ associated to the partition $n+1=n_1+\cdots+n_k$ can be expressed as
$$
J_{+}^{(n_k,\ldots,n_1)} = \sum_{j=1\atop n_j\not=1}^{k} \>
\left(\left(b^{(\rm p)}_{1}\right)_{A_{n_j-1}} \right)_0\>,
\nfr{Connect}
where $(\;)_0$ indicates the projection of the components whose
homogeneous grade is zero.

The eqs.\Gradator\ and \Connect\ specify some sort of canonical
relation between the non-conjugated embeddings of $A_1$ into $A_n$ and
the non-equivalent Heisenberg subalgebras of $A_{n}^{(1)}$, a relation which
is possible to generalize for the other classical simple Lie
algebras~[\Ref{CARL}]. Nevertheless, it only involves the
particular elements of the Heisenberg subalgebra directly related to the
affine cyclic element of Kostant. But, in general, one needs to relate the
zero-grade component of any element of an arbitrary Heisenberg
subalgebra to the $A_1$ subalgebras. In the case of $A_{n}^{(1)}$ it is
possible to do it by using the explicit representation
\PrinHeis. Let us take $n\geq2$ and $l=1,\ldots,n-1$, then $n=ml+r$ for some
$0\leq r<l$, where $m$ is the integer part of the ratio $n/l$. Next,
consider the nilpotent element $\left(b_{l}^{({\rm p})}\right)_0\in
A_{n-1}$ as the
$J_+$ of an $A_1$ subalgebra of $A_{n-1}$. After a tedious calculation,
one can show that $J_0$ is the  matrix~[\Ref{HAM2},\Ref{FEHFRAC}]
$$
J_0 = {\rm diag} \biggl(\underbrace{m,\ldots,m}_{r{\rm\; times}} \>,\>
\underbrace{m-1,\ldots,m-1}_{(l-r){\rm\; times}}\>,\>\ldots \>,\>
\underbrace{-m,\ldots,-m}_{r{\rm\; times}}\biggr)\>,
\nfr{Fractional}
where the multiplicities $r$ and $(l-r)$ occur alternately, ending
always with $r$. Let us point out that, within this representation of
$A_{n-1}$, the main diagonal of $J_0$ is precisely the defining vector
of the embedding; therefore, using \Embdef, one can prove the following
result.

\jump\noindent
{\bf Lemma 8.} {\it The nilpotent element, $\left(b_{l}^{({\rm
p})}\right)_0\in A_{n-1}$, where $n=ml+r\geq2$ with $l\geq1$, $0\leq
r<l$, and $m\in{\Bbb Z}>0$, can be considered as the $J_+$ of the $A_1$
subalgebra of $A_{n-1}$ associated to the partition
$$
n\> =\> \underbrace{(m+1)+\cdots+(m+1)}_{r{\rm\; times}}\>
+\> \underbrace{m+\cdots+m}_{(l-r){\rm\; times}}\>,
\nfr{PartFrac}
\ie, to the regular subalgebra
$$
\left(A_{m}\right)^{\oplus r}\oplus \left(A_{m-1}\right)^{\oplus(l-r)}
\subset A_{n-1}\>.\qquad
\square
\nfr{FractPlus}
}\jump

This result leads to the generalization of \Connect\ in an obvious way, but
we shall omit the explicit formula because it becomes too cumbersome.
In any case, this shows that there are different forms of expressing
the $J_+$ of an $A_1$ subalgebra of $A_n$ as the zero graded component
of some element of a Heisenberg subalgebra of $A_{n}^{(1)}$.

The following result of~[\Ref{FEHFRAC}] will be used in the next
section.

\jump\noindent
{\bf Lemma 9.} {\it Let $B_-$ be the maximal nilpotent negative
subalgebra of $A_{n-1}$; \ie, it corresponds to the lower triangular
$n\times n$ matrices in the fundamental representation of
$sl(n,{\Bbb C})$. Then
$$
{\rm Ker}\left(\ad \bigl(b_{l}^{({\rm p})}\bigr)_0\right)\> \cap \>B_-
=\emptyset
\efr
if and only if either $l=1$, or $l=2$ and $n$ is odd. $\square$}\jump

Finally, let us mention the following result of~[\Ref{FHM}] that is
relevant in relation to the distinction between hierarchies of type~I
and type~II ---recently, this result has been generalized for the other
classical Lie algebras in~[\Ref{DFEH}].

\jump\noindent
{\bf Proposition 4.} (Theorem~1.7 of [\Ref{FHM}]) {\it Graded regular
elements exist only in those Heisenberg subalgebras of $A_{n}^{(1)}$
which belong to the special partitions
$$
n+1= \underbrace{a+\cdots +a}_{p{\rm\; times}}
{\it\>\;\; and\>\;\;} n+1=
\underbrace{a+\cdots +a}_{p{\rm\; times}}+1\>,
$$
where $a>1$. The graded regular elements are of the form
$$
\Lambda = z^m \left( \sum_{j=1}^{p}\> y_j\> \left(b_{l}^{({\rm p})}
\right)_{A_{a-1}}\right)\>,
\nfr{Regular}
where
$$
1\leq l\leq a-1\>,\qquad y_i\not=0\>,\qquad y_{i}^{a} \not=
y_{j}^{a}\>,\quad
i,j=1,\ldots,p\>, \quad i\not=j\>,
$$
$l$ is relatively prime to $a$, and $m\in {\Bbb Z}$.
When
$$
n+1= \underbrace{a+\cdots +a}_{p{\rm\; times}}+1
$$
and $a$ is even, the element
$\Lambda$ has grade $2(l+ma)$ with respect to the gradation \Gradation, where
$N_{\sw}=2a$. Otherwise, $\Lambda$ is of grade $(l + ma)$ and $N_{\sw}=a$.}
$\square$\jump

\chapter{{\bf \W-algebras and integrable hierarchies of the KdV type.}}

We are now in position to investigate which are the
\W-algebras arising as the second Poisson bracket algebra of the
integrable hierarchies of~[\Ref{GEN1},\Ref{GEN2}] in the sense of theorem~2.
In the following, we restrict ourselves to the generalized KdV hierarchies
based on $A_{n}^{(1)}$, for which $\s$ is the homogeneous gradation
$\s_{\rm h}=(1,0,\ldots,0)$.

Theorem~2 indicates that, under
\Degenplus, the second Poisson bracket algebra of the integrable hierarchy
associated to the data $\lbrace \gg, \Heis,\sw, \s,
\Lambda\rbrace$ contains the \W-algebra characterized by $J_+=
(\Lambda)_0$. Therefore, it will be possible to associate at least one
integrable hierarchy to those \W-algebras corresponding to the $A_1$
subalgebras of $A_n$ such that $J_+$ can be expressed as the zero
homogeneous grade component of an element of some Heisenberg subalgebra
with definite
$\sw$-grade. The results of the previous section allow us to express
the $J_+$ of any $A_1$ subalgebra of $A_n$ as the zero homogeneous grade
component of, at least, one element of some Heisenberg subalgebra.
Accordingly, the condition will be that at least one of these elements have
definite $\sw$-grade.

Let us analyse first the canonical connection expressed by \Connect.
With respect to the gradation $\sw$ induced by \Gradator, the cyclic
element
$\left(b^{(K)}\right)_{A_{n_j-1}}$ has grade $N_{\sw}/n_j$. Consequently,
the condition that the element of the Heisenberg subalgebra on the
right-hand-side of \Connect\ has definite $\sw$-grade requires that all
the $n_j\not=1$ are equal. This restricts the choice of $J_+$ to
partitions of the type
$$
n+1 = \underbrace{a+\>\cdots+\>a}_{p{\;\rm times}} +
\underbrace{1+\>\cdots+\>1}_{q{\;\rm times}} \equiv p\>(a) + q\>(1)\>,
\nfr{Partplus}
with $a>1$. With this restriction,
$$
H_{\sw}\> =\> {N_\sw\over 2a}\> J_{0}^{(a,\ldots,a,1,\ldots,1)}\>,
\nfr{Gradplus}
where
$$
[w]=[\> \underbrace{a,\ldots,a}_{p{\;\rm times}} \>,\>
\underbrace{1,\ldots,1}_{q{\;\rm times}}\>]\>.
\nfr{Conjclass}
Notice that $H_{\sw}$ is proportional to the $J_0$ of the
relevant $A_1$ subalgebra, which ensures that condition \Degenplus\ is indeed
satisfied ---see comments below theorem~2. Moreover, according to
lemma~4, the $\sw$-grade of $\Lambda$, $i$, is either $1$ or $2$.

The gradations associated to the partitions \Partplus\ are the following.

\jump\noindent
{\bf Lemma 10.} {\it The gradation $\sw$ corresponding to the conjugacy class
\Conjclass\ is
$$
\sw=\bigl(\underbrace{1,0_{p-1},\ldots,1,0_{p-1}}_{{a+1\over2}{\;\rm
times}}\>,\> 0_q\>,\>
\underbrace{1,0_{p-1},\ldots,1,0_{p-1}}_{{a-1\over2}{\;\rm times}}\bigr)\>,
\nfr{Gradodd}
with $N_{\sw}= a$, when $a$ is an odd number,
$$
\sw=\bigl(\underbrace{2,0_{p-1},\ldots,2,0_{p-1}}_{{a\over2}{\;\rm times}}\>,\>
1,0_{q-1},1,0_{p-1}\>,\>\underbrace{2,0_{p-1},
\ldots,2,0_{p-1}}_{{a\over2}-1{\;\rm times}}\bigr)\>,
\nfr{Gradeven}
with $N_{\sw}= 2a$, when $a$ is even and $q\not=0$, and
$$
\sw=\bigl(\underbrace{1,0_{p-1},\ldots,1,0_{p-1}}_{a {\;\rm
times}}\bigr)\>,
\nfr{Gradeveno}
with $N_{\sw}= a$, when $a$ is even and $q=0$. In these equations
we have used the notation $0_j=\underbrace{0,\ldots,0}_{j{\;\rm times}}$}.

\sjump\noindent
{\it Proof.} When $a$ is odd, it follows from \Double\ that $N_{\sw}=
a$. Besides, \Gradator\ leads to
$$
\eqalign{
H_{\sw}& \equiv (\gamma_1,\ldots,\gamma_{n+1})\cr
&= \left(\bigl({a-1\over
2}\bigr)_p,\bigl({a-3\over 2}\bigr)_p, \ldots
,1_p,0_{p+q},-1_p,\ldots,-\bigl({a-1\over 2}\bigr)_p\right)
\>,\cr}
\efr
where we have already ordered the components of $H_\sw$ as indicated below
\Pregrad. Using \Gradation, all this leads to \Gradodd. In a similar way,
when $a$ is even, $N_{\sw}= 2a$, and
$$
H_{\sw} \equiv (\gamma_1,\ldots,\gamma_{n+1})=
\left((a-1)_p,(a-3)_p,\ldots,1_p,0_{q},-1_p,\ldots,-(a-1)_p\right)\>,
\efr
which proves \Gradeven\ and \Gradeveno. $\square$\sjump

Let us indicate that the $\sw$-grade of the element of $\Heis$ that appears
on the right-hand-side of \Connect\ for the partition \Partplus\ is
$i=({\sw})_0$, \ie, either $i=1$ or $i=2$ according to lemma~10; this is the
lowest positive $\sw$-grade of the elements of $\Heis$.

So far, we have only analysed the canonical relation \Connect\ between the
$A_1$ subalgebras of $A_n$ and the Heisenberg subalgebras of $A_{n}^{(1)}$.
We can conclude that it is only possible to derive the $\cal
W$-algebras associated to the partitions of the form \Partplus\ from the
second Poisson bracket algebras constructed in terms of the Heisenberg
subalgebras associated to the same partitions. Nevertheless, as we have
already said in the previous section, there are more general
connections between $A_1$ subalgebras and Heisenberg subalgebras that follow
from lemma~8. Let us consider the conjugacy class
$$
[w]=[\> n_1,\ldots,n_k,\>, \underbrace{1,\ldots,1}_{q{\rm\; times}} \>]
\nfr{GenConj}
of the Weyl group of $A_n$, with $n_j\not=1$ for $j=1,\ldots,k$, and the
nilpotent element
$$
J_+ = \left( \sum_{j=1}^{k} y_j \left(b_{l_j}^{({\rm
p})}\right)_{A_{n_j-1}} \right)_0\>,
\nfr{General}
with $1\leq l_j<n_j$ for $j=1,\ldots,k$. In this case, the condition that the
element of $\Heis$ defining the right-hand-side of \General\ has
definite
$\sw$-grade is accomplished if
$$
{N_{\sw}\over n_1}l_1 = {N_{\sw}\over n_2}l_2 =\cdots {N_{\sw}\over n_k}l_k\>
\efr
which implies that
$$
n_j = m\>l_j + r_j\>, \qquad j=1,\ldots,k\>,
\nfr{Gradao}
with $m\in{\Bbb Z}\geq1$, $0\leq r_j<l_j$, and either $r_j=0$ for all
$j=1,\ldots,k$, or $r_j\not=0$ and $r_i/r_j =l_i/l_j$ for any $i,j=1,
\ldots,k$. Hence, it follows from lemma~8 that the
$J_+$ expressed by \General\ corresponds to the embedding of $A_1$ associated
to
the partition
$$
n=\underbrace{(m+1)+ \cdots+(m+1)}_{\sum_{j=1}^{k} r_j{\;\rm times}}\>+\>
\underbrace{m+ \cdots+m}_{\sum_{j=1}^{k}(l_j- r_j){\;\rm times}}\>+\>
\underbrace{1+\cdots+1}_{q{\;\rm times}}\>.
\nfr{GenPart}

Let us impose now the condition \Degenplus. Our starting point is the
observation that, because of eq.\Gradator, the restriction
$$
H_{\sw}\biggm|_{A_{n_j-1}} \> =\> {N_{\sw}\over 2n_j}\> \left(J_{0}^{({\rm
p})}\right)_{A_{n_j-1}}\>,
\efr
which implies that
$$
\bigoplus_{j=1\atop n_j\not=1}^{k}\> (B_-)_{A_{n_j-1}}\> \subsetneq\>
P\>= \>\gg_{0}(\s) \cap \gg_{<0}(\sw)\>,
\nfr{Nilpotent}
where $(B_-)_{A_{n_j-1}}$ is the maximal negative nilpotent
subalgebra of $A_{n_j-1}$. Therefore, lemma~9 can be applied to each
factor $\left(b_{l_j}^{({\rm p})}\right)_{A_{n_j-1}}$ in eq.\General, which
requires that\note{We thank Laszlo Feh\'er for pointing
out this fact to us~[\Ref{LAZLO}].}
$$
{\rm either}\quad
l_j= 1\>, \quad{\rm or}\quad l_j = 2 \quad{\rm and}\quad n_j \quad{\rm
is\;\; odd\;\; for\;\; all\;\;} j=1,\ldots,k\>.
\nfr{Fehcond}
In general, because the left-hand-side of eq.\Nilpotent\ does not equal $P$,
\Fehcond\ is just a necessary condition to ensure that \Degenplus\ is
satisfied. Nevertheless, when combined with the condition that the element of
$\Heis$ on the right-hand-side of \General\ has definite
$\sw$-grade, it implies that all the $n_j$'s have to be equal; then,
according to \Gradator, $H_{\sw}$ is proportional to the $J_0$ of the $A_1$
subalgebra corresponding to $J_+$ in \General, and, consequently the condition
\Degenplus\ is indeed satisfied in all these cases, as we have discussed
after theorem~2,.

In conclusion, eqs.\Gradao, \GenPart, and \Fehcond\ specify the only $\cal
W$-algebras that can be derived as the second Poisson bracket algebra of the
integrable hierarchies of~[\Ref{GEN1},\Ref{GEN2}] based on a constant
element $\Lambda\in\Heis$ that satisfies the condition \Degenplus. This,
together with theorem~2, is the main result of this paper and we summarize
it as follows.

\jump\noindent
{\bf Theorem 3.} {\it (i) Graded elements that satisfy condition \Degenplus\
exist only in those Heisenberg subalgebras of $A_{n}^{(1)}$ associated to
the special conjugacy classes
$$
[w] =[\underbrace{m,\ldots,m}_{k {\rm\; times}}\>,
\underbrace{1,\ldots,1\/}_{q {\rm\; times}}]\>.
\efr
These elements are of the form
$$
\Lambda^{(1)}= \sum_{j=1}^{k}\> y_j\> \left( b_{1}^{({\rm
p})}\right)_{A_{m-1}}
\>,
\efr
for any value of $m$, and of the form
$$
\Lambda^{(2)} = \sum_{j=1}^{k}\> {y}_{j}\>\left(  b_{2}^{({\rm p})}
\right)_{A_{m-1}}\>,
\efr
when $m$ is odd. If either $m$ is odd or $m$ is even and $q=0$, the
$\sw$-grade of $\Lambda^{(1)}$ equals $1$; otherwise, $\Lambda^{(1)}\in
\gg_{2}(\sw)$. In any case, the $\sw$-grade of $\Lambda^{(1)}$ is the lowest
positive $\sw$-grade of the elements of $\Heis$. The $\sw$-grade of
$\Lambda^{(2)}$ equals $2$.
\newline (ii) Consider now the integrable hierarchies of the KdV type
associated to
$A_{n}^{(1)}$, $\Heis$, $\sw$, $\s_{\rm h}=(1,0,\ldots,0)$, and to one of
those two elements of $\Heis$. The algebras defined by the second Poisson
bracket of these integrable hierarchies, restricted to the gauge invariant
functionals of $q_0(x)$, are the $\cal W$ algebras associated to the
embeddings of $A_1$ into $A_n$ labelled by the partitions
$$
n=\underbrace{m+ \cdots+m}_{ k {\;\rm times}}\>
+\> \underbrace{1+\cdots+1}_{q{\;\rm times}} \equiv k(m) + q(1)\>,
\nfr{Patauno}
for $\Lambda^{(1)}$, or
$$
\eqalign{
n &=\underbrace{(a+1)+ \cdots+(a+1)}_{k {\;\rm times}}\>
+\>\underbrace{a+ \cdots+a}_{k {\;\rm times}}\>
+\> \underbrace{1+\cdots+1}_{q{\;\rm times}}\cr
&\equiv k(a+1) + k(a) + q(1) \>, \cr}
\nfr{Patados}
for $\Lambda^{(2)}$ when $m$ is odd, $m=2a+1$. $\square$}

\sjump\noindent
{\it Remark.} Using proposition~4, one can restrict theorem~3 to the
integrable hierarchies of type~I. Then, $q$ has to be either $0$ or
$1$, and $y_{i}^{m}\not= y_{j}^{m}$ for any
$i\not=j=1,\ldots,k$. \jump

Notice that the connection expressed by theorem~3 shows that the
\W-algebras specified by the partitions $n= \>k(2)\> +\> q(1)$
are shared as second Poisson bracket algebras by the integrable hierarches
associated to the conjugacy classes
$$
[w]=[\underbrace{2,\ldots,2}_{k{\rm\;times}}\>,
\underbrace{1,\ldots,1}_{q{\rm\;times}}]\>, {\rm\;\; with\;\;}
\Lambda^{(1)}\in \Heis\>,
\efr
and
$$
[w]=[\underbrace{3,\ldots,3}_{k{\rm\;times}}\>,
\underbrace{1,\ldots,1}_{q-k{\rm\;times}}]\>, {\rm\;\; with\;\;}
\Lambda^{(2)}\in \Heis\>.
\efr
Nevertheless, in general, the \W-algebras associated to the partitions
\Patauno\ or \Patados\ can be obtained only from integrable hierarchies where
$\Lambda$ has to be of the form $\Lambda^{(1)}$ or $\Lambda^{(2)}$,
respectively.

Finally, let us point out that the class of \W-algebras covered by
theorems~2 and~3 is very restricted even if type~II hierarchies are
considered; a fact that has been already anticipated in~[\Ref{FHM}]. In
fact, notice that the structure of the second Poisson bracket algebra of
the integrable hierarchies is actually independent of the fact that they are
of type~I or of type~II as far as \Degen\ is satisfied. The first
$\cal W$-algebra that cannot be recovered as the second Poisson bracket
algebra of any of the integrable hierarchies of~[\Ref{GEN1},\Ref{GEN2}], with
$\Lambda$ restricted by the condition \Degenplus, corresponds to the partition
$6=4+2$,\ie, to the regular subalgebra $A_3\oplus A_1\subset A_5$.

\chapter{{\bf Examples.}}

\section{The First Fractional $A_2$  KdV-Hierarchy and  ${\cal
W}_{3}^{(2)}$ again.}

This paper has been largely motivated by the analysis of this
case in~[\Ref{GEN2}]; here, we discuss it again to illustrate some points of
the previous sections. Let us consider the KdV hierarchy corresponding
to the principal Heisenberg subalgebra of
$A_{2}^{(1)}$ and to the element $\Lambda$ with principal grade $i=2$,
$$
\Lambda = \pmatrix{0 & 0 & 1\cr
                   z & 0 & 0\cr
                   0 & z & 0\cr}\>.
\nfr{LambFrac}
Since this $\Lambda$ is not the element of the
principal Heisenberg subalgebra with minimal positive grade,
this integrable hierarchy can be considered just as the result of
changing $x$ by the first time of the generalized $A_2$ Drinfel'd-Sokolov
hierarchy~[\Ref{MATH1},\Ref{MATH2}]. Before gauge fixing, the potential can
be written as
$$
q(x) = \pmatrix{y_1(x) & c(x) & 0\cr
                e(x) & y_2(x) & d(x)\cr
                a(x) + z\, b(x) & f(x) & -(y_1(x) + y_2(x))\cr}\,
\nfr{PotFrac}
which, under a gauge transformation, becomes
$$
q(x) \mapsto \Phi(x)\partial_x\Phi^{-1}(x)  + \Phi(x)\left(\Lambda +
q(x)
\right)\Phi^{-1}(x) - \Lambda\>,
$$
where
$$
\Phi(x) = \exp S(x)= \pmatrix{1 & 0 & 0\cr
                   A(x) & 1 & 0\cr
                   B(x) & C(x) & 1\cr}\>.
\efr
First of all, let us identify the $A_1$ subalgebra
of $A_2$ corresponding to this case, whose generators are
$$
J_+= \left(\Lambda\right)_0 =\pmatrix{0 & 0 & 1\cr
                                      0 & 0 & 0\cr
                                      0 & 0 & 0\cr}\>,\quad
J_0= \pmatrix{1 & 0 & 0\cr
              0 & 0 & 0\cr
              0 & 0 & -1\cr}\>, {\rm\>\; and\;\>}
J_-= \pmatrix{0 & 0 & 0\cr
              0 & 0 & 0\cr
              1 & 0 & 0\cr}\>.
\efr
According to Section~5, the form of $J_0$ indicates that it is the
$A_1$ subalgebra corresponding to the partition $3=2+1$. Moreover,
the principal gradation is induced precisely by $H_{\sw} =
J_0$ which obviously fulfills the condition \Degenplus. Now,
instead of  fixing the gauge as in~[\Ref{GEN1},\Ref{GEN2}], we choose the
gauge slice according to theorem~1. Then,
$$
q^{({\rm can})}(x) = 3\phi(x)\pmatrix{0 & 0 & 0\cr
                                      0 & 0 & 0\cr
                                      z & 0 & 0\cr}
+\pmatrix{U(x)/2 & 0 & 0\cr
          G^+(x) & -U(x) & 0\cr
          T(x) & G^-(x) & U(x)/2 \cr}\>,
\efr
which leads to the functional dependence on $q_0(x)$ of the components of
$\widehat{S}(x)$ (or equivalently $\widehat{\Phi}(x)$)
$$
\widehat{A} = -d \>,\quad \widehat{C} = c\>, \quad
\widehat{B} = {1\over2}\left( -c d + 2y_1 + y_2\right)\>,
\efr
and of the components of ${\cal J}(x)$
$$
\eqalign{
&\phi = {1\over3}\left( b + c + d\right)\cr
&U = c d - y_2\cr
&G^+  = - c d^2 + e - d\left( y_1- y_2\right) +
d'\cr
&G^-  = - c^2 d + f + c\left( y_1 + 2y_2\right) -
c'  \>.\cr}
\efr
The energy-momentum tensor \Tfinal\ is
$$
\eqalign{
{\cal T}(x) &= T \> +\> {3\over4} U^2\cr
&= a + c e + d\>f + y_{1}^{2} + y_1y_2 + y_{2}^{2}\cr
&\qquad + {1\over2}\left( cd' - d c'\right) -
y_{1}' -{1\over2} y_{2}'\>.\cr}
\efr
One can check that $\phi$, $U$, $G^{\pm}$, and ${\cal T}$
are exactly equal to the combinations $\phi$, $\tilde U$, $\tilde
G^{\pm}$, and $\tilde T$ of~[\Ref{GEN2}], respectively. We already knew
that, in this case, the restriction of the second Poisson bracket to
the components of $q^{({\rm can})}_0(x)$ is a \W-algebra; in fact, it
is the $\cal W$-algebra associated to the $A_1$ subalgebra of $A_2$
labelled by the partition $3=2+1$. This realization of this \W-algebra
is an example  of the connection expressed by theorem~3, with
$[w]=[3]$ and $\Lambda=\Lambda^{(2)}$.

But it also follows from theorem~3 that this \W-algebra can also
be realized as the second Poisson bracket of a different integrable
hierarchy with $[w]=[2,1]$ and $\Lambda= \Lambda^{(1)}$. This is the canonical
connection expressed by \Connect, \Partplus, and \Conjclass, which leads
precisely to the realization discussed in~[\Ref{GEN2}] (see
also~[\Ref{VAND}]), with $J_+= (\Lambda')_0$ and
$$
\Lambda' = \Lambda^{(1)} = \pmatrix{0 & 0 & 1\cr
                   0 & 0 & 0\cr
                   z & 0 & 0\cr} \in {\cal H}[2,1]\>.
\efr
Finally, let us recall that
$\Lambda$ and $\Lambda'$ are regular elements of the corresponding
Heisenberg subalgebras of $A_{2}^{(1)}$ and, hence, these integrable
hierarchies are of type~I.

\section{Other Fractional $A_{N-1}$ KdV-Hierarchies and \W-algebras.}

The example analysed in the previous section is just the simplest
case of the so called fractional KdV-hierarchies~[\Ref{BAKF}]. Let us
now consider the KdV  general hierarchy corresponding to the
principal Heisenberg subalgebra of $A_{N-1}^{(1)}$, $[w]=[N]$, and to
the element
$\Lambda$ with principal grade $i=2$, \ie,
$$
\Lambda= b_{2}^{({\rm p})} =\pmatrix{0 & {\Bbb I}_{(N-2)\times(N-2)}\cr
                  z\> {\Bbb I}_{2\times 2}& 0\cr}\>.
\efr
According to lemma~8, the embedding of $A_1$ into $A_{N-1}$
corresponding to $J_+= \left( b_{2}^{({\rm p})} \right)_0$ is labelled
by the partition
$$
N= \left({N+1\over 2}\right)\> +\> \left({N-1\over 2}\right)
\nfr{PartOdd}
when $N\geq3$ is odd, and by the partition
$$
N= \left({N\over 2}\right)\> +\> \left({N\over 2}\right)
\nfr{PartEven}
when $N\geq4$ is even. Nevertheless, before claiming any relation between the
second Poisson bracket of this hierarchy and \W-algebras, one has to
check if the condition \Degenplus\ is satisfied.

Let us start with the case
when $N$ is odd. Then, it follows from lemma~9 that indeed
the condition \Degenplus\ holds in this case. Then, the second Poisson
bracket algebra of this KdV hierarchy contains the \W-algebra
associated to the partition \PartOdd, which is nothing else than the
fractional ${\cal W}_{N}^{(2)}$ algebra of~[\Ref{BER},\Ref{FEHFRAC}].
This case is the straightforward generalization of the first example
that we have analysed in this section, which is recovered when $N=3$.

Let us now study the case when $N$ is even. Then, it is easy to check
that
$$
\theta= \sum_{j=1}^{N/2} E_{2j,2j-1} \in P= g_0(\s)\cap g_{<0}(\sw)
\efr
belongs to ${\rm Ker}(\ad J_+)$ as well; hence, the condition
\Degenplus\ is not satisfied in agreement with lemma~9. Even more, one can
check that $\bigl[\Lambda, \theta\bigr]=0$; therefore, neither \Degenplus\
nor \Degen\  are satisfied, and one cannot construct an integrable
hierarchy in the sense of eq.\Gaugeflow\ in this case.\jump

Let us now go on to consider the KdV hierarchy corresponding to the
element $\Lambda$ with principal grade $i=3$, \ie,
$$
\Lambda= b_{3}^{({\rm p})} =\pmatrix{0 & {\Bbb I}_{(N-3)\times(N-3)}\cr
                  z\> {\Bbb I}_{3\times 3}& 0\cr}\>.
\efr
The $A_1$ subalgebra of $A_N$ with $J_+=\left( b_{2}^{({\rm p})}
\right)_0$ is labelled by the partition
$$
N= \left({N+2\over 3}\right)\> +\> \left({N-1\over 3}\right)
+\> \left({N-1\over 3}\right)
\nfr{Partto}
when $N\in 3{\Bbb Z}+1\geq4$, by the partition
$$
N= \left({N+1\over 3}\right)\> +\> \left({N+1\over 3}\right)\>
+\>\left({N-2\over 3}\right)\>,
\nfr{Parttt}
when $N\in3{\Bbb Z}+2\geq5$, and by the partition
$$
N= \left({N\over 3}\right)\> +\> \left({N\over 3}\right)
+\> \left({N\over 3}\right)\>,
\nfr{Parttz}
when $N\in3{\Bbb Z} \geq6$.

It is easy to check that
$$
\theta_1= \sum_{j=1}^{(N-1)/3} E_{3j,3j-1} \in P
\efr
belongs to ${\rm Ker}(\ad J_+)$ when $N\in 3{\Bbb Z}+1\geq4$, that
$$
\theta_2= \sum_{j=1}^{(N+1)/3} E_{3j-1,3j-2} \in P
\efr
belongs to ${\rm Ker}(\ad J_+)$ when $N\in3{\Bbb Z}+2\geq5$, and that
$$
\theta_3= \sum_{j=1}^{N/3}\left( \alpha E_{3j,3j-1} + \beta
E_{3j-1,3j-2}+ \gamma E_{3j,3j-2}\right) \in P
\efr
belongs to ${\rm Ker}(\ad J_+)$ when $N\in3{\Bbb Z} \geq6$ for arbitrary
constants $\alpha$, $\beta$, and $\gamma$. Thus, we conclude that the
condition \Degenplus\ is not satisfied in this case, as anticipated by
lemma~9. Nevertheless, one can notice that $[\Lambda,\theta_3]=0$, but
$[\Lambda,\theta_1]$ and
$[\Lambda,\theta_2]$ do not vanish. Hence, \Degen\ holds except when
$N\in3{\Bbb Z} \geq6$ and one can indeed construct integrable
hierarchies in all the other cases. It is expected that their second
Poisson bracket algebras will contain \W-algebras as well, but they are
not described by the results of this
paper~(see~[\Ref{MATH2},\Ref{BAKDEP}]) .

Finally, let us point out that all these fractional $A_{N-1}$
hierarchies can be described as the result of changing
$x$ for some of the times $t_{b}$ of the generalized $A_{N-1}$
Drinfel'd-Sokolov hierarchy, in  the sense
of~[\Ref{MATH1},\Ref{MATH2}], only when $b\in\Heis$ is regular and the
resulting hierarchy is of type~I.

\section{A Fractional KdV-Hierarchy associated to ${\cal H}[3,3]\subset
A_{5}^{(1)}$.}

The last example is an integrable hierarchy such that
$q^{({\rm can})}_{>0}(x)$ has dynamical components, \ie, it has some
components that are not centres of the second Poisson bracket. Within
the $A_{n}^{(1)}$ algebras, the simplest case is associated to the
Heisenberg subalgebra of $A_{5}^{(1)}$ corresponding to the conjugacy
class of the Weyl group $[w]=[3,3]$.

According to \Double, $N_{\sw}=3$, and, using \Gradator, one gets
(see the appendix),
$$
H_{\sw} = {\rm diag\>}\left( {3\over 2\cdot3}(2,0,-2) \cup {3\over
2\cdot3}(2,0,-2) \right) \mapsto {\rm diag\>}(1,1,0,0,-1,-1)\>,
\efr
and the gradation associated to ${\cal H}[3,3]$ is
$$
\sw = (1,0,1,0,1,0)\> \>.
\efr
Using the basis of simple roots associated to $\sw$, the generators
of the Heisenberg subalgebra are
$$
b_{3m} = z^m\> {\rm diag\>}(1,-1,1,-1,1,-1)
\efr

$$
b_{1+3m}^{(1)} = z^m\>  \pmatrix{& &0& & & \cr
                         & & &1& & \cr
                         & & & &0& \cr
                         & & & & &1\cr
                        0& & & & & \cr
                         &z& & & & \cr}\>\qquad
b_{1+3m}^{(2)} = z^m\>  \pmatrix{& &1& & & \cr
                         & & &0& & \cr
                         & & & &1& \cr
                         & & & & &0\cr
                        z& & & & & \cr
                         &0& & & & \cr}
\efr

$$
b_{2+3m}^{(1)} = z^m\>  \pmatrix{& & & &0& \cr
                         & & & & &1\cr
                        0& & & & & \cr
                         &z& & & & \cr
                         & &0& & & \cr
                         & & &z& & \cr}\>\qquad
b_{2+3m}^{(2)} = z^m\>  \pmatrix{& & & &1& \cr
                         & & & & &0\cr
                        z& & & & & \cr
                         &0& & & & \cr
                         & &z& & & \cr
                         & & &0& & \cr}\>,
\efr
where $m\in{\Bbb Z}$ and the subscripts are the $\sw$-grades.

Let us consider the type~I hierarchy associated to
$A_{5}^{(1)}$, ${\cal H}[3,3]$, $\s_{[3,3]}$, $\s_{\rm h}=(1,0,\ldots,0)$,
and  the regular element
$$
\Lambda=\pmatrix{& & & &1& \cr
                         & & & & &2\cr
                        z& & & & & \cr
                         &2\>z& & & & \cr
                         & &z& & & \cr
                         & & &2\>z& & \cr}\> =\> b_{2}^{(2)}\>
+\>2 \>b_{2}^{(1)} \in {\cal H}[3,3]\>,
\efr
whose grade is $i=2$. In block form, the potential is
$$
q(x)= \pmatrix{{\Bbb A}(x)&{\Bbb B}(x)&0\cr
               {\Bbb C}(x)&{\Bbb D}(x)&{\Bbb E}(x)\cr
               {\Bbb F}(x)&{\Bbb G}(x)&{\Bbb H}(x)\cr} +
z\>\pmatrix{0&0&0\cr
            0&0&0\cr
            {\Bbb J}(x)&0&0\cr}\in Q\>,
\efr
where all the entries are $2\times 2$ matrices, and the gauge
transformations are generated by elements of the form
$$
S(x)=\pmatrix{0&0&0\cr
              {\Bbb T}(x)&0&0\cr
              {\Bbb U}(x)&{\Bbb V}(x)&0\cr}\in P\>;
\efr
$q(x)$ has $35$ independent components, while $S(x)$ has $12$.
According to theorem~1, the gauge slice can be chosen such that
$$
\eqalignno{
q^{({\rm can})}(x) &= \pmatrix{a_1&a_2&0&0&0&0\cr
2a_3&-a_4&0&0&0&0\cr
b_1&b_2&a_5-2a_1&a_6&0&0\cr
b_3&b_4&a_7&2a_4-a_5&0&0\cr
c_1&c_2&b_5&b_6&a_1&2a_2\cr
c_3&c_4&b_7&b_8&a_3&-a_4\cr}\>\cr
\noalign{\vskip 0.3cm}
& +\>
z\> \pmatrix{0&0&0&0&0&0\cr 0&0&0&0&0&0\cr 0&0&0&0&0&0\cr
0&0&0&0&0&0\cr \phi&B&0&0&0&0\cr
A&\chi&0&0&0&0\cr}\>,&\numali \cr}
$$
\ie, it has $23(=35-12)$ independent components.

It follows from theorem~3, with $[w]=[3,3]$ and $\Lambda=\Lambda^{(2)}$,
\ie, with $m=3$, $k=2$, and $q=0$, that $\{a_1,\ldots,a_7, b_1,\ldots,b_8,
c_1,\ldots,c_4\}$ generate the \W-algebra associated to the embedding
of $A_1$ into $A_5$ labelled by the partition $6= (2)+(2)+(1)+(1)$. The
conformal structure is defined by
$$
{\cal T}(x) = c_1 + 2c_4 + 3(a_{1}^{2} + a_{4}^{2}) + a_{5}^{2} -
2(a_1 + a_4) a_5 + 4 a_2 a_3 + a_6 a_7\>,
\efr
and the generators $\{a_1,\ldots,a_7\}$, $\{b_1,\ldots,b_8\}$, and
$\{c_1-2c_4,c_2,c_3\}$ have conformal spin $1$, $3/2$, and $2$,
respectively. Moreover, in analogy with the case of
${\cal W}_{3}^{(2)}$, notice that this
\W-algebra could also be obtained from another integrable hierarchy
associated to the Heisenberg subalgebra
${\cal H}[2,2,1,1]$ and to $\Lambda=\Lambda^{(1)}$.

In contrast, the second Poisson brackets of the components of
$q^{({\rm can})}_{1}(x)$ are not given by theorems~2 and~3. It is easy
to check that $\phi(x)$ and
$\chi(x)$ are the two centres associated by proposition~1 to
$b_{1}^{(1)}$ and $b_{1}^{(2)}$, and, hence, they are not dynamical
quantities. The brackets involving the other two components, $A(x)$ and
$B(x)$, can be calculated with the help of the computer. The result
is that the only non-zero second Poisson bracket is
$$
\{A(x)\> ,\>B(y)\}_2\> =\> {7\over2}\> \delta(x-y)\>,
\efr
and we conclude that $A(x)$ and $B(x)$ generate a
{\it ``b--c''} algebra that is decoupled from the \W-algebra generated
by the components of $q^{({\rm can})}_{0}(x)$. Therefore, in this case,
it is actually possible to write down an energy-momentum tensor that
generates the conformal transformation of all the components of
$q^{({\rm can})}(x)$
$$
\widehat{{\cal T}}(x) = {\cal T}(x) +{1\over7}\biggl( A(x)B'(x) -
A'(x)B(x)\biggr)\>,
\efr
which gives conformal spin $1\over2$ to $A(x)$ and $B(x)$.

\chapter{{\bf Conclusions and discussion.}}

We have presented a systematic study of the second Hamiltonian
structures of the integrable hierarchies defined
in~[\Ref{GEN1},\Ref{GEN2}] in terms of (untwisted) affine Kac-Moody
algebras. They are based on a graded element
$\Lambda$ of some Heisenberg subalgebra, and we have restricted
ourselves to the case when ${\rm Ker} ({\rm ad}\> (\Lambda)_0)
\cap \gg_{<0}(\sw)=\emptyset$. When
$(\Lambda)_0\not=0$, the second Poisson bracket algebra contains one of the
(classical)
$\cal W$-algebras defined through the method of Drinfel'd-Sokolov
Hamiltonian reduction in [\Ref{EMB},\Ref{HAM1},\Ref{HAM2}]. We have
also generalized some of the results of~[\Ref{GEN1},\Ref{GEN2}] for the
type~II hierarchies, but, even considering both type~I and type~II
hierarchies, the class of $\cal W$-algebras that can be obtained in
this way is very limited. The restriction comes both from the condition
that $\Lambda$ is a {\it graded} element and from the condition that it
satisfies ${\rm Ker} ({\rm ad}\>  (\Lambda)_0) \cap \gg_{<0}(\sw)=\emptyset$.
It implies that, in the case of $A_{n}^{(1)}$ algebras, only the $\cal
W$-algebras based on the embeddings of $A_1$ into $A_n$ labelled by the
partitions of the form
$$
n+1 \> =\> k\>(m)\> +\> q\>(1)\quad {\rm and}\quad n+1\> =\> k\>(m+1)\> +\>
k\>(m) \>+\> q\>(1)
$$
can be defined as the second Poisson bracket algebras of some of these
integrable hierarchies. A relevant feature is that some of these
$\cal W$-algebras can be recovered from more
than one integrable hierarchy. Notice that there
is a well defined ``Miura map'' within each integrable hierarchy, an
that the Miura map usually leads to the realization of the $\cal
W$-algebra in terms of free fields~[\Ref{FL}]. Therefore, there might
be more than one possible realizations suggested by this connection.

Let us point out that our results cover most of the $\cal W$-algebras
that have already been related to integrable hierarchies of
the KdV type in the literature. An example of the other ones are the
${\cal W}_{4}^{(2)}$ and ${\cal W}_{4}^{(3)}$ algebras
of~[\Ref{BAKDEP}] and~[\Ref{MATH2}], respectively. The problem is that,
in these cases, either the condition \Degen\ is not satisfied, or it is
not possible to restrict the second Poisson bracket algebra to
functions that only depend on
$q_0(x)$. We think that a further generalization of our results to
cover these cases can be achieved by comparing directly the second
Poisson bracket with the Dirac bracket of the Hamiltonian reduction;
this work is already in progress. More recently, a series of  $\cal
W$-algebras associated to a new generalization of the Gel'fand and
Dickey's pseudo-differential approach has been constructed in
ref.[\Ref{BON}]. The relation of these $\cal W$-algebras to the
Drinfel'd-Sokolov Hamiltonian reduction and, of course, to our approach
is unclear yet. It would be very interesting to understand better this
connection not only from the point of view of the classification of
$\cal W$-algebras, but also in order to have and alternative definition
of the integrable hierarchies of~[\Ref{GEN1},\Ref{GEN2}]. In fact, it
has already been shown in~[\Ref{FHM},\Ref{DFEH}] that some of them can be
defined using a matrix version of the he Gel'fand--Dickey's
pseudo-differential approach.

Finally, let us comment briefly about the possible applications of our
results to the study of two-dimensional integrable field theories.  It
has already been said in the introduction that the $\cal W$-algebras
are naturally related to the (conformal) Toda field theories.
Therefore, our results should help to a better understanding of the
relationship between those conformal Toda field theories and the
integrable hierarchies of the KdV type. Some steps in this direction
has already been taken in ref.[\Ref{FRENK}]. Even more, it is possible
to associate a generalization of the Toda equation to the equations of
the mKdV type~[\Ref{DS},\Ref{WIL}]. In some cases, this generalized
Toda equation can be understood as the equation of motion of a massive
affine Toda field theory, and all this could clarify the connection
between conformal and massive Toda models. In relation to this, we
expect that our detailed discussion of the $A_1$ subalgebras of a
simple Lie algebra and the Heisenberg subalgebras of its untwisted
affinization will be particularly useful~[\Ref{UNDER}].

\appendix{{\bf Appendix.}}

We present the non-equivalent
Heisenberg subalgebras of $A_{2}^{(1)}$ and $A_{3}^{(1)}$ (see
also~[\Ref{FHM}] and~[\Ref{KROOD1}]). We shall realize $A_n$ in  terms
of the fundamental representation of $sl(n+1,{\Bbb C})$; then, the
basis \Basis\ is just
$$
H_i = E_{i,i}\>,\qquad E_{(e_j-e_k)} = E_{k,j}\>,
\efr
where $E_{j,k}$ is the $(n+1)\times (n+1)$ matrix whose only
non-vanishing entry is a $1$ at the position $(j,k)$. Here, we do not
mention neither the principal Heisenberg subalgebra, associated to
$[n+1]$, nor the homogeneous Heisenberg subalgebra, associated to
$[1,\ldots,1]$.

\sjump\noindent
${\bf A_2}:$

\sjump\noindent
\item{$\bullet$} $[w]= [2,1]$ ($A_1\subset A_2$): $N_{\sw}=4$, and
$$
H_{\sw} = {\rm diag\>} \left( {4\over 2\cdot 2}(1,-1)\cup(0)\right)
\mapsto {\rm diag\>}(1,0,-1)\>,
\nfr{Apone}
where the arrow indicates that we have ordered the components of
$H_{\sw}$; in this case, $\gamma_1 \mapsto \gamma_1$, $\gamma_2 \mapsto
\gamma_3$, and $\gamma_3 \mapsto \gamma_2$. Then,
$$
{\sw} = (2,1,1)\>, \quad{\rm and}\quad g_0({\sw}) = u(1)^2\>.
\efr
The generators of the Heisenberg subalgebra are
$$
\eqalignno{&b_{4m} = z^m \pmatrix{ 1& & \cr
                        &1& \cr
                        & &-2\cr} \mapsto
z^m \pmatrix{ 1& & \cr
               &-2& \cr
               & &1\cr}\>,\cr
\noalign{\vskip 0.3cm}
&b_{2+ 4m} = z^m \pmatrix{0&1& \cr
                      z&0& \cr
                       & &0\cr} \mapsto
z^m \pmatrix{ & &1\cr
          &0& \cr
         z& & \cr}\>, \qquad m\in{\Bbb Z}\>,&\numali\cr}
$$
where the arrow indicates the same permutation of components as in
\Apone, and the subscripts are the grades with respect to $\sw$. As we
have explained in the main text, this permutation corresponds to a
change of the basis of the simple roots through conjugation with the Weyl
group, to ensure that $H_{\sw}$ indeed induces a ${\Bbb Z}$-gradation. Notice
that the elements of $\Heis$ have block-diagonal form before performing
this permutation.\sjump

\item{}To illustrate the non-uniqueness of the gradation $\sw$, let
us point out that the generators $b_{4m}$ and $b_{2+4m}$ have definite
grade with respect to any gradation of the form $\sw'=m(s_1
+s_2,s_1,s_2)$ their grades being
$2m(s_1+s_2)$ and $(1+2m)(s_1+s_2)$, respectively.

\sjump\noindent
${\bf A_3}:$

\sjump\noindent
\item{$\bullet$} $[w]= [3,1]$ ($A_2\subset A_3$): $N_{\sw}=3$, and
$$
H_{\sw} = {\rm diag\>} \left( {3\over 2\cdot 3}(2,0,-2)\cup(0)\right)
\mapsto {\rm diag\>}(1,0,0,-1)\>;
\efr
then,
$$
{\sw} = (1,1,0,1)\>, \quad{\rm and}\quad g_0({\sw}) =  A_1\oplus
u(1)^2\>.
\efr
The generators of the Heisenberg subalgebra are
$$
\eqalignno{
&b_{3m} = z^m \pmatrix{1& & & \cr
                        &1& & \cr
                        & &1& \cr
                        & & &-3\cr} \mapsto
z^m \pmatrix{1& & & \cr
              &1& & \cr
              & &-3& \cr
              & & &1\cr}\>, \cr
\noalign{\vskip 0.3cm}
&b_{1+ 3m} = z^m \pmatrix{0&1&0& \cr
                          0&0&1& \cr
                          z&0&0& \cr
                           & & &0\cr} \mapsto
z^m \pmatrix{0&1&0&0\cr
             0&0&0&1\cr
             0&0&0&0\cr
             z&0&0&0\cr}\>, \cr
\noalign{\vskip 0.3cm}
&b_{2+ 3m} = z^m \pmatrix{0&0&1& \cr
                          z&0&0& \cr
                          0&z&0& \cr
                           & & &0\cr} \mapsto
z^m \pmatrix{0&0&0&1\cr
             z&0&0&0\cr
             0&0&0&0\cr
             0&z&0&0\cr}\>. &\numali\cr}
$$

\sjump\noindent
\item{$\bullet$} $[w]= [2,2]$ ($A_1\oplus A_1\subset A_3$):
$N_{\sw}=2$,
$$
H_{\sw} = {\rm diag\>} \left( {2\over 2\cdot 2}(1,-1)\cup{2\over 2\cdot
2}(1,-1)\right)\mapsto {\rm diag\>}({1\over2},{1\over2} , -{1\over2},
-{1\over2})\>,
\efr
$$
{\sw} = (1,0,1,0)\>, \qquad g_0({\sw}) = A_1\oplus A_1\oplus
u(1)\>,
\efr
and
$$
\eqalignno{
&b_{2m} = z^m \pmatrix{1& & & \cr &1& & \cr
                       & &-1& \cr & & &-1\cr} \mapsto
z^m \pmatrix{1& & & \cr &-1& & \cr
                       & &1& \cr & & &-1\cr}\>, \cr
\noalign{\vskip 0.3cm}
&b_{1+2m}^{(1)} = z^m \pmatrix{0&1& & \cr z&0& & \cr
                             & &0& \cr  & & &0\cr}\mapsto
z^m \pmatrix{0&0&1&0\cr 0&0&0&0\cr
                            z&0&0&0\cr 0&0&0&0\cr}\>, \cr
\noalign{\vskip 0.3cm}
&b_{1+2m}^{(2)} = z^m \pmatrix{0& & & \cr  &0& & \cr
                             & &0&1\cr  & &z&0\cr}\mapsto
z^m \pmatrix{0&0&0&0\cr 0&0&0&1\cr
                              0&0&0&0\cr 0&z&0&0\cr}\>,
&\numali\cr}
$$
where the superscripts distinguish the different generators of
$\Heis$ having the same $\sw$-grade.

\sjump\noindent
\item{$\bullet$} $[w]= [2,1,1]$ ($A_1\subset A_3$): $N_{\sw}=4$,
$$
H_{\sw} = {\rm diag\>} \left( {4\over 2\cdot
2}(1,-1)\cup(0)\cup(0)\right)\mapsto {\rm diag\>}(1,0,0,-1)\>,
\efr
$$
{\sw} = (2,1,0,1)\>, \qquad g_0({\sw}) = A_1\oplus u(1)^{2}\>,
\efr
and
$$
\eqalignno{
&b_{4m}^{(1)} = z^m \pmatrix{1& & & \cr &1& & \cr
                       & &-2& \cr & & &0\cr} \mapsto
z^m \pmatrix{1& & & \cr &0& & \cr
                       & &-2& \cr & & &1\cr}\>, \cr
\noalign{\vskip 0.3cm}
&b_{4m}^{(2)} = z^m \pmatrix{0& & & \cr &0& & \cr
                       & &-1& \cr & & &1\cr} \mapsto
z^m \pmatrix{0& & & \cr &1& & \cr
                       & &-1& \cr & & &0\cr}\>, \cr
\noalign{\vskip 0.3cm}
&b_{2+4m} = z^m \pmatrix{0&1& & \cr z&0& & \cr
                             & &0& \cr  & & &0\cr}\mapsto
z^m \pmatrix{0&0&0&1\cr 0&0&0&0\cr
                            0&0&0&0\cr z&0&0&0\cr}\>. &\numali\cr}
$$
\sjump

For $A_4$ and $A_5$ we just show the gradations $\sw$ and the grades of the
elements of the corresponding Heisenberg subalgebras. First, for the
conjugacy classes of the Weyl group of $A_4$,
$$
\vbox{\settabs 3 \columns
\+{\bf Conjugacy} & {\bf Gradation
$\sw$} & {\bf Grades of $\Heis$} \cr
\+{\bf class $[w]$} &  &  &  \cr
\+ $[4,1]$ & $(2,2,1,1,2)$ & $(0,2,4,6)$ mod $8$ \cr
\+ $[3,2]$ & $(4,1,3,3,1)$ & $(0,4,6,8)$ mod $12$\cr
\+ $[3,1,1]$ & $(1,1,0,0,1)$ & $(0,0,1,2)$ mod $3$\cr
\+ $[2,2,1]$ & $(2,0,1,1,0)$ & $(0,0,2,2)$ mod $4$\cr
\+ $[2,1,1,1]$ & $(2,1,0,0,1)$ & $(0,0,0,2)$ mod $4$\cr}
$$
and, second, for the Weyl group of $A_5$,
$$
\vbox{\settabs 3 \columns
\+{\bf Conjugacy} & {\bf Gradation
$\sw$} & {\bf Grades of $\Heis$} \cr
\+{\bf class $[w]$} &  &  &  \cr
\+ $[5,1]$ & $(1,1,1,0,1,1)$ & $(0,1,2,3,4)$ mod $5$\cr
\+ $[4,2]$ & $(2,1,1,2,1,1)$ & $(0,2,4,4,6)$ mod $8$\cr
\+ $[4,1,1]$ & $(2,2,1,0,1,2)$ & $(0,0,2,4,6)$ mod $8$\cr
\+ $[3,3]$ & $(1,0,1,0,1,0)$ & $(0,1,1,2,2)$ mod $3$\cr
\+ $[3,2,1]$ & $(4,1,3,0,3,1)$ & $(0,0,4,6,8)$ mod $12$\cr
\+ $[3,1,1,1]$ & $(1,1,0,0,0,1)$ & $(0,0,0,1,2)$ mod $3$\cr
\+ $[2,2,2]$ & $(1,0,0,1,0,0)$ & $(0,0,1,1,1)$ mod $2$\cr
\+ $[2,2,1,1]$ & $(2,0,1,0,1,0)$ & $(0,0,0,2,2)$ mod $4$\cr
\+ $[2,1,1,1,1,]$ & $(2,1,0,0,0,1)$ & $(0,0,0,0,2)$ mod $4$\cr}
$$

%
%

\jump\jump
\centerline{\bf Acknowledgements}

\sjump\noindent
We would like to thank Tim Hollowood, Luiz Ferreira and Philippe
Ruelle for their very useful comments and discussions. We are specially
indebted to Laszlo Feh\'er for having suggested the use of lemma~9 in
section~6. The research reported in this paper has been supported partially
by C.I.C.Y.T (AEN93-0729) and D.G.I.C.Y.T. (PB90-0772). CRFP and MVG are
also supported partly by Xunta de Galicia.

\jump\jump
\references

\beginref

\Rref{REV}{P.~Bouwknegt and K.~Schoutens, Phys. Rep. {\bf 223} (1993)
183; \newline P.~van Driel, {\sl Construction and Applications of
Extended Conformal Symmetries\/}, PhD thesis, Univ. of Amsterdam (1992);
\newline T.~Tjin, {\sl Finite and Infinite W Algebras and their
Applications\/}, PhD thesis, Univ. of Amsterdam (1993).}
\Rref{ZAM}{A.B.~Zamolodchikov, Theor. Math. Phys. {\bf 65} (1985) 347.}
\Rref{GD}{I.M.~Gel'fand and L.A.~Dikii, Funkts. Anal. Pril. {\bf 10}
(1976) 13; Funkts. Anal. Pril. {\bf 13} (1979) 13.}
\Rref{DS}{V.G.~Drinfel'd and V.V.~Sokolov, J.~Sov. Math. {\bf 30}
(1985) 1975; Soviet. Math. Dokl. {\bf 23} (1981) 457.}
\Rref{FL}{S.L.~Luk'yanov and V.A.~Fateev, Sov. Sci. Rev. A. Phys. {\bf 15}
(1990) 1.}
\Rref{BAIS}{F.A.~Bais, P.~Bouwknegt, K.~Schoutens, and M.~Surridge,
Nucl. Phys. {\bf B304} (1988) 348-370; Nucl. Phys. {\bf B304} (1988) 371-391.}
\Rref{BAK}{I.~Bakas, Commun. Math. Phys. {\bf 123} (1989)
627; Phys. Lett. {\bf B213} (1988) 313;\newline
P.~Mathieu, Phys. Lett. {\bf B208} (1988) 101.}
\Rref{HAM1}{J.~Balog, L.~Feh\'er, P.~Forg\'acs, L.~O'Raifeartaigh, and
A.~Wipf, Ann. Phys. (N.Y.) {\bf 203} (1990) 76.}
\Rref{HAM2}{L.~Feh\'er, L.~O'Raifeartaigh, P.~Ruelle,
I.~Tsutsui, and A.~Wipf, Phys. Rep. {\bf 222} (1992) 1.}
\Rref{EMB}{F.A.~Bais, T.~Tjin, and P.~van Driel, Nucl. Phys. {\bf B357}
(1991) 632;\newline L.~Feh\'er, L.~O'Raifeartaigh, P.~Ruelle, I.~Tsutsui,
and A.~Wipf, Ann. Phys. (N.Y.) {\bf 213} (1992) 1.}
\Rref{RUELLE}{L.~Feh\'er, L.~O'Raifeartaigh, P.~Ruelle, and I.~Tsutsui,
Commun. Math. Phys. {\bf 162} (1994) 399.}
\Rref{CASES}{T.~Tjin and P.~van Driel, {\sl Coupled WZNW-Toda Models and
Covariant KdV Hierarchies\/}, preprint ITFA-91-04;\newline
F.~Toppan, Phys. Lett. {\bf B327} (1994) 249.}
\Rref{FHM}{L.~Feh\'er, J.~Harnad, and I.~Marshall, Commun.
Math. Phys. {\bf 154} (1993) 181;\newline
L.~Feh\'er, {\sl Generalized Drinfeld-Sokolov Hierarchies and \W-algebras},
Proceedings of the NSERC-CAP Workshop on Quantum Groups, Integrable Models
and Statistical Systems, Kingston, Canada, 1992.}
\Rref{GEN1}{M.F.~de Groot, T.J.~Hollowood, and J.L.~Miramontes, Commun. Math.
Phys. {\bf 145} (1992) 57.}
\Rref{GEN2}{N.J.~Burroughs, M.F.~de Groot, T.J.~Hollowood, and
J.L.~Miramontes, Commun. Math. Phys. {\bf 153} (1993)
187; Phys. Lett. {\bf B277} (1992) 89.}
\Rref{NIGEL1}{N.J.~Burroughs, Nonlinearity {\bf 6} (1993) 583.}
\Rref{NIGEL2}{N.J.~Burroughs, Nucl. Phys. {\bf B379} (1992) 340.}
\Rref{TAU}{T.J.~Hollowood, and J.L.~Miramontes, Commun. Math. Phys. {\bf 157}
(1993) 99.}
\Rref{KW}{V.G.~Kac and M.~Wakimoto, Proceedings of Symposia in Pure
Mathematics, vol. {\bf 49} (1989) Part I, 191.}
\Rref{KOST}{B.~Kostant, Amer. J. Math. {\bf 81} (1959) 973.}
\Rref{UNDER}{J.~Underwood, {\sl Aspects of Non-Abelian Toda Theories\/},
preprint Imperial/TP/92 -93/30, hep-th 9304156.}
\Rref{FRENK}{B.~Feigin and E.~Frenkel, {\sl Kac-Moody Groups and
Integrability of Soliton Equations\/}, preprint RIMS-970 (1994).}
\Rref{KP}{V.G.~Kac and D.H.~Peterson, Symposium on Anomalies,
Geometry and Topology, W.A.~Bardeen and A.R.~White (eds.), Singapore, World
Scientific (1985) 276-298.}
\Rref{JAC}{N.~Jacobson, {\sl Lie Algebras\/}, Wiley-Interscience, New York
(1962).}
\Rref{SORB1}{L.~Frappat, E.~Ragoucy and P.~Sorba, Commun. Math. Phys. {\bf
157} (1993) 499.}
\Rref{SORB2}{F.~Delduc, E.~Ragoucy and P.~Sorba, Phys. Lett. {\bf B279}
(1992) 319.}
\Rref{KACB}{V.G.~Kac, {\sl Infinite Dimensional Lie Algebras ($3^{\rm rd}$
ed.)\/}, Cambridge University Press, Cambridge (1990).}
\Rref{CARELK}{R.W.~Carter and G.B.~Elkington, J. of Algebra {\bf 20} (1972)
350.}
\Rref{DYN}{E.B.~Dynkin, Amer. Math. Soc. Trans. Ser.~2 {\bf 6} (1957)
111.}
\Rref{LG}{M.~Lorente and B.~Gruber, J. Math. Phys. {\bf 13} (1972)
1639.}
\Rref{RAG}{E.~Ragoucy, Nucl. Phys. {\bf B409} (1993) 213-236.}
\Rref{CART}{R.W.~Carter, Compositio Mathematica {\bf 25} Fasc.~1 (1972) 1.}
\Rref{KROOD1}{F.~ten Kroode and J.~van de Leur, Commun. Math. Phys.
{\bf 137} (1991) 67.}
\Rref{KROOD2}{F.~ten Kroode and J.~van de Leur, Commun. in
Algebra {\bf 20} (1992) 3119; Acta Appl. Math. {\bf 27} (1992) 153; Acta Appl.
Math. {\bf 31} (1993) 1.}
\Rref{CARL}{C.R.~Fern\'andez-Pousa and J.L.~Miramontes, in preparation.}
\Rref{MATH1}{P.~Mathieu and W.~Oevel, Mod. Phys. Lett. {\bf A6}
(1991) 2397.}
\Rref{MATH2}{D.A.~Depireux and P.~Mathieu, Int. J. of Mod. Phys. {\bf A7}
(1992) 6053.}
\Rref{VAND}{P.~van Driel, Phys. Lett. {\bf 274B} (1991) 179.}
\Rref{BAKF}{I.~Bakas and D.A.~Depireux,  Mod. Phys. Lett. {\bf A6}
(1991) 1561, ERRATUM ibid. {\bf A6} (1991) 2351; Int. J. Mod. Phys.
{\bf A7} (1992) 1767.}
\Rref{BER}{M.~Bershadsky, Commun. Math. Phys. {\bf 139} (1991) 71;\newline
A.~Polyakov, Int. J. Mod. Phys. {\bf A5} (1990) 833.}
\Rref{FEHFRAC}{L.~Feh\'er, L.~O'Raifeartaigh, P.~Ruelle, and I.~Tsutsui,
Phys. Lett. {\bf B283} (1992) 243.}
\Rref{LAZLO}{L.~Feh\'er, private communication.}
\Rref{BAKDEP}{I.~Bakas and D.A.~Depireux, {\sl Self-Duality, KdV Flows and
W-Algebras\/}, Proceedings of the $XX^{th}$ International Conference on
Differential Geometric Methods in Theoretical Physics, New York, June 1991.}
\Rref{BON}{L.~Bonora and C.S.~Xiong, {\sl The $(N,M)^{th}$ KdV Hierarchy and
the associated W Algebras\/}, SISSA 171/93/EP, BONN-HE/45/93, hep-th
9311070;\newline L.~Bonora, Q.P.~Liu, and C.S.~Xiong, {\sl The Integrable
Hierarchy constructed from a pair of higher KdV Hierarchies and its
associated W Algebra\/}, BONN-TH-94-17, hep-th 9408035.}
\Rref{WIL}{G.W.~Wilson, Ergod. Theor. \& Dyn. Sist. {\bf 1} (1981) 361.}
\Rref{DFEH}{F.~Delduc and L.~Feh\'er, {\sl Conjugacy Classes in the Weyl
Group Admitting a Regular Eigenvector and Integrable Hierarchies\/},
ENSLAPP-L-493/94, hep-th 9410203.}

\endref
\ciao